\documentclass[sigconf, nonacm]{acmart}
\setcopyright{rightsretained}

\renewcommand\footnotetextcopyrightpermission[1]{}

\newcommand\vldbdoi{XX.XX/XXX.XX}
\newcommand\vldbpages{XXX-XXX}
\newcommand\vldbvolume{16}
\newcommand\vldbissue{10}
\newcommand\vldbyear{2023}
\newcommand\vldbauthors{\authors}
\newcommand\vldbtitle{\shorttitle} 

\newcommand\vldbpagestyle{empty} 
\usepackage{etoolbox}
\newtoggle{fullversion}
\toggletrue{fullversion}
\usepackage{amsmath}

\usepackage{url}
\usepackage{makecell}
\usepackage{scalerel}
\usepackage{balance}
\usepackage{epstopdf}
\usepackage{adjustbox}
\usepackage{graphicx}
\usepackage{subfigure}
\usepackage{color}
\usepackage{hhline}
\usepackage{tabularx}
\usepackage{hyperref}
\usepackage{ragged2e}  
\newcolumntype{Y}{>{\RaggedRight\arraybackslash}X}
\usepackage{booktabs}
\usepackage{fixltx2e}
\usepackage{paralist}
\usepackage{listings}
\usepackage{multirow}
\usepackage{enumerate}
\usepackage{bbm}
\usepackage{xspace}
\usepackage{comment}
\usepackage[inline]{aplcomments}
\usepackage{caption} 

\usepackage[noend,lined,boxed,vlined,ruled,linesnumbered]{algorithm2e}
\SetKwInOut{Input}{input}
\SetKwInOut{Output}{output}

\usepackage{amsmath}
\DeclareMathOperator*{\argmax}{arg\!\max}
\DeclareMathOperator*{\argmin}{arg\!\min}

\usepackage{comment}
\usepackage[labelfont=bf]{caption}
\usepackage{fixltx2e}
\usepackage{enumitem}
\usepackage{etoolbox} 
\usepackage{mdwlist} 

\usepackage{array}
\newcolumntype{L}{>{\centering\arraybackslash}m{2cm}}

\newcommand{\abi}{\textsc{Auto-BI}\xspace}
\newcommand{\pbi}{\textsc{System-X}\xspace}

\newcommand{\code}[1]{\texttt{#1}}

\newcounter{definition}
\newenvironment{definition}[1][]{\refstepcounter{definition}\par\smallskip\textsc{Definition~\thedefinition.\ #1}}{\smallskip}

\newcounter{example}
\newenvironment{example}[1][]{\refstepcounter{example}\par\smallskip\textsc{Example~\theexample.\ #1}}{\smallskip}

\newcounter{theorem}
\newenvironment{theorem}[1][]{\refstepcounter{theorem}\par\smallskip\textsc{Theorem~\thetheorem.\ #1}}{\smallskip}

\newcounter{proposition}

\makeatletter
  \newcommand\figcaption{\def\@captype{figure}\caption}
  \newcommand\tabcaption{\def\@captype{table}\caption}
\makeatother

\begin{document}

\iftoggle{fullversion}{\setlength{\floatsep}{0pt}
\setlength{\textfloatsep}{0pt}
\setlength{\abovecaptionskip}{0pt}
\setlength{\abovedisplayskip}{0pt}
\setlength{\belowdisplayskip}{0pt}
\setlength{\itemsep}{0pt}
\setlength{\partopsep}{0pt}}{
\setlength{\floatsep}{0pt}
\setlength{\textfloatsep}{0pt}
\setlength{\abovecaptionskip}{0pt}
\setlength{\abovedisplayskip}{0pt}
\setlength{\belowdisplayskip}{0pt}
\setlength{\itemsep}{0pt}
\setlength{\partopsep}{0pt}
}

\pagenumbering{gobble}


\title{Auto-BI: Automatically Build BI-Models \\ Leveraging   Local Join Prediction and Global Schema Graph}

\author{Yiming Lin}
\authornote{Work done at Microsoft.}
\affiliation{%
  \institution{University of California, Irvine}
}
\email{yiminl18@uci.edu}

\author{Yeye He}
\affiliation{%
  \institution{Microsoft Research}
}
\email{yeyehe@microsoft.com}

\author{Surajit Chaudhuri}
\affiliation{%
  \institution{Microsoft Research}
}
\email{surajitc@microsoft.com}

\begin{abstract}
Business Intelligence (BI) is crucial in modern enterprises and billion-dollar business. Traditionally, technical experts like database administrators would manually prepare BI-models (e.g., in star or snowflake schemas) that join tables in data warehouses, before less-technical business users can run analytics using end-user dashboarding tools.
However, the popularity of self-service BI (e.g., Tableau and Power-BI) in recent years creates a strong demand for less technical end-users to build BI-models themselves. 

We develop an \abi system that can accurately predict BI models given a set of input tables, 
using a principled graph-based optimization problem we propose called \textit{k-Min-Cost-Arborescence} (k-MCA), which holistically considers both local join prediction and global schema-graph structures, leveraging a graph-theoretical structure called \textit{arborescence}. While we prove k-MCA is intractable and inapproximate in general, we develop novel algorithms that can solve k-MCA optimally, which is shown to be efficient in practice with sub-second latency and can scale to the largest BI-models we encounter (with close to 100 tables). 

\abi is rigorously evaluated on a unique dataset with over 100K real BI models we harvested, as well as on 4 popular TPC benchmarks. It is shown to be both efficient and accurate, achieving over 0.9 F1-score on both real and synthetic benchmarks.
\end{abstract}

\maketitle

\pagestyle{\vldbpagestyle}
\begingroup\small\noindent\raggedright\textbf{PVLDB Reference Format:}\\
\vldbauthors. \vldbtitle. PVLDB, \vldbvolume(\vldbissue): \vldbpages, \vldbyear.\\
\href{https://doi.org/\vldbdoi}{doi:\vldbdoi}
\endgroup
\begingroup
\renewcommand\thefootnote{}\footnote{\noindent
This work is licensed under the Creative Commons BY-NC-ND 4.0 International License. Visit \url{https://creativecommons.org/licenses/by-nc-nd/4.0/} to view a copy of this license. For any use beyond those covered by this license, obtain permission by emailing \href{mailto:info@vldb.org}{info@vldb.org}. Copyright is held by the owner/author(s). Publication rights licensed to the VLDB Endowment. \\
\raggedright Proceedings of the VLDB Endowment, Vol. \vldbvolume, No. \vldbissue\ %
ISSN 2150-8097. \\
\href{https://doi.org/\vldbdoi}{doi:\vldbdoi} \\
}\addtocounter{footnote}{-1}\endgroup


\vspace{-2mm}
\section{Introduction}

Business Intelligence (BI) is increasingly important in modern enterprises for data-driven decision making, and has grown into a multi-billion dollar business~\cite{gartner-bi}.
In traditional BI settings, database administrators (DBAs) typically need to manually prepare BI-models (table schemas and join relationships) in data warehouses, so that less-technical business users can perform ad-hoc analysis using tools like dashboards~\cite{chaudhuri2011overview}. 

In recent years, in a growing trend called ``\textit{self-service BI}''~\cite{ss-bi} that is popularized by vendors like Tableau~\cite{tableau} and Power-BI~\cite{power-bi}, less-technical business users are increasingly expected to set up and perform BI analysis themselves, without relying on DBAs or central IT. The goal is to democratize BI, so that business users can make agile data-driven decisions themselves, without depending on technical users.

\textbf{Building BI models: still a pain point.} At a high level, there are two main steps in BI project: (1) building BI models, and (2) performing ad-hoc analysis by querying against BI models. While querying BI-models was made simple by vendors like Tableau and Power-BI (through intuitive user interfaces and dashboards)~\cite{mackinlay2007show, hanrahan2006vizql}, the first step of building ``BI-models'', a prerequisite for ad-hoc analysis, remains a key pain point for non-technical users. 

\begin{figure*}[t]
	\centering
\vspace{-15mm}
	\includegraphics[width=1\linewidth]{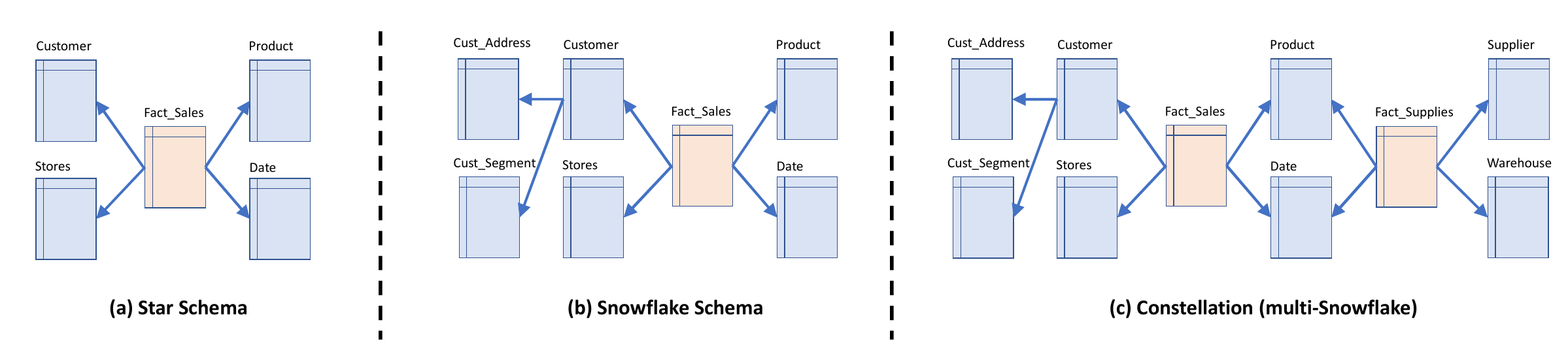}
\vspace{-4mm}
	\caption{Types of common BI schemas. (a) Star schema, which can be seen as a simplified version of Snowflake. (b) Snowflake schema, where dimension tables like ``Customers'' can connect to additional dimension tables. (c) Constellation schema, which can be seen as multiple Snowflake schemas.}
	\vspace{-1em}
	\label{fig:schemas}
\end{figure*}

In the context of self-service BI tools like Tableau and Power-BI, ``BI-modeling'' refers to the process  of preparing raw data and establishing relationships between tables, where a central task is to establish join relationships for a given set of input tables\footnote{We note that more advanced BI modeling can involve additional steps such as schema redesign and performance optimization, which is usually out of scope for non-technical users in self-service BI tools, and thus not considered in this work.}. This closely relates to foreign-key (FK) detection~\cite{hpi, chen2014fast, zhang2010multi} but works specifically in the context of BI, where the resulting schema graphs from the modeling step frequently correspond to structures known as star-schema and snowflake-schema studied in data warehouses~\cite{chaudhuri2011overview}, like shown in Figure~\ref{fig:schemas}.

While self-service BI tools also attempt to improve the usability of the BI-modeling step through better GUI
 (e.g., allowing users to specify join columns using drag-and-drop)~\cite{pq-drap-drop-model, tableau-drap-drop-model}, building BI models remains a key pain point. This is because when faced with a large number of tables, even experienced technical users like DBAs can find the task of identifying all possible join relationships challenging and time-consuming. For less-technical enterprise users who are not familiar with concepts like fact/dimension tables, building BI models from scratch can be a daunting challenge.

\textbf{Foreign-key detection: not yet sufficient.}
It would clearly be useful, if the join relationships in BI models can be automatically predicted on given input tables (without requiring users  to specify them manually).  Since joins in BI models are often primary-key (PK) foreign-key (FK) joins,  existing FK detection algorithms~\cite{hpi, chen2014fast, zhang2010multi} would seem to apply.

To study this systematically, we harvested over 100K real BI models built using self-service BI tools,
from  public sources like GitHub and search engines. For each BI model file, we programmatically extract the input tables used in the model, as well as the ground-truth join relationships  specified by users, thus creating  a  real BI benchmark for large-scale evaluation for the first time (prior work  mostly use synthetic benchmarks for evaluation instead). 

Our large-scale evaluation using these real BI datasets suggests that existing FK-detection algorithms are still insufficient for the task, as they frequently produce incorrect results (with \textasciitilde 0.6 precision for all methods we tested). This is because real BI data in the wild tend to be considerably more challenging (e.g., column-headers can often be cryptic with generic names like ``name'' or ``id'', and column-values can often overlap accidentally). Crucially, most existing approaches are ``local'' in nature, that consider two tables at a time to make greedy local decisions, without a principled global optimization, thus  unable to produce accurate predictions on challenging real BI test cases (Section~\ref{sec:related}).



\textbf{\abi: global optimization using schema graphs.} 
A key insight we develop in \abi, is that because we know users are finding join relationships specifically for building BI-models, and we know the typical structure that schema-graphs in BI-models  should generally follow (e.g., variants of snowflake), this gives us a unique opportunity to leverage the \textit{global graph structure} of the resulting schema-graph, to predict joins much more accurately.

We thus formulate the \abi problem as a global optimization problem that holistically considers both the \textit{local decision} of pair-wise joinability, as well as the \textit{global decision} of graph structures. 

Specifically, we show that the snowflake schema popular in BI-modeling corresponds to 
a graph-theoretical concept called \textit{Arborescence}~\cite{rosen2008its}. 
Leveraging this structure, we formulate a new graph problem called 
\textit{k-MCA (k-Minimum-Cost Arborescence)}, which finds the most probable $k$-snowflakes, by considering both local joinability and  global graph structures, all using a  precise probabilistic interpretation based on calibrated  probabilities.




We prove that the new $k$-MCA problem and its variants are in general intractable (NP-hard) and inapproximable (EXP-APX-complete). Nevertheless, we develop novel algorithms based on branch-and-bound principles, which are surprisingly efficient on real test cases and can scale to the largest BI models we encounter, while still being provably optimal.
Our extensive evaluations using real BI models and synthetic TPC benchmarks suggest that the proposed \abi is  substantially more accurate when compared to existing solutions in the literature. 

\textbf{Contributions.} We make the following contributions:

\begin{itemize}[noitemsep,topsep=0pt,leftmargin=*]
\item We are the first to harvest and leverage over 100K real BI models in the wild, for the problem of predicting BI models. This enables a data-driven algorithm design and makes it possible to rigorously evaluate different algorithms on real BI data. 
\item We are the first to exploit the snowflake-like schema structure in BI settings for our predictions, by mapping snowflakes to a less-known graph-theoretical concept called arboresence. We formulate a set of novel graph-based optimization problems that we call $k$-MCA, that have precise probabilistic foundations.
\item We study the theoretical hardness of $k$-MCA variants, and propose efficient algorithms that are provably optimal. Extensive evaluations show that our algorithms are both effective (with over 0.9 F1 scores), and efficient (scales to the largest BI models we encounter and runs in a few seconds).
\end{itemize}

\vspace{-2mm}
\section{Related Works}
\label{sec:related}
\textbf{BI and dashboarding tools.}
There are a wide variety of BI and dashboarding tools that aim to help users perform ad-hoc data analysis, with Tableau~\cite{tableau} and Power-BI~\cite{power-bi} being the leading vendors~\cite{gartner-bi}. These tools use visual drag-and-drop interfaces~\cite{mackinlay2007show} (without requiring users to write SQL), and are particularly popular among non-technical users.

\textbf{Foreign key detection.} Foreign key (FK) detection is an important problem in database settings, with many influential methods developed in the literature~\cite{zhang2010multi, chen2014fast, hpi, fk-ml}.

MC-FK~\cite{zhang2010multi} is a pioneering effort to detect FK using an EMD-based randomness metric for distribution similarity between two columns, which is more reliable to predict true relationships when unrelated key-columns that can frequently have overlapping ranges.

Fast-FK~\cite{chen2014fast} develops an efficient method that selects FKs with the best pre-defined scores until all input tables are connected.

HoPF~\cite{hpi} improves upon prior work by considering not only FK-scores but also PK-scores, which are combined using a predefined scoring function, making this also a global method in spirit. The algorithm enumerates all PK/FK combinations and returns the combination that has the highest total score.

ML-FK~\cite{fk-ml} proposes an ML-based classifier to predict FK, trained on known FKs. As we will show analytically and experimentally, without principled global optimizations, ML classification alone is still local in nature and not sufficient to achieve high accuracy. 

While FK detection methods are clearly related to our problem (we experimentally compare them with these methods), there are also a few key distinctions that we would like to highlight.

First, while FK detection targets general database settings, we focus on predicting joins in BI models, which gives us a unique opportunity to exploit the likely graph structure (e.g., snowflake) to make more accurate predictions, which is a unique direction not considered in prior work.

Second, unlike FK-detection methods that typically consider the canonical PK/FK (1:N) joins,  in \abi the types of joins we consider are more general, because real BI models in the wild frequently employ 1:1 joins as well as joins that are not perfectly 1:N, making the prediction problem more challenging.

Lastly, with the exception of~\cite{fk-ml}, most prior FK detection methods primarily rely on hand-crafted scoring functions to make local join predictions (pairwise between two tables). In comparison, we leverage the BI models harvested to first predict local-joinability in a data-driven way, which is then combined into a principled global optimization for optimal global decisions (at the graph-level for all tables). This also makes our technique different from prior work.


\textbf{Detect inclusion dependency.} 
Inclusion dependency (IND) is closely related to FK, with an influential body of work focusing on  \textit{efficiently} enumerating inclusion dependency (IND) in large databases~\cite{ind-1, ind-2, ind-3, ind-4, ind-5, ind-6, ind-7, ind-8}. The focus on efficiency makes this line of work orthogonal to FK-detection, where the focus is to accurately predict meaningful FKs (from a large collection of IND candidates). Like prior FK-detection methods that employ efficient IND detection~\cite{hpi, zhang2010multi}, we also use IND-detection as a pre-processing step to enumerate possible FK candidates efficiently.

\textbf{Complex non-equi joins.}
Beyond equi-joins, techniques  to automatically detect and handle complex join relationships have also been studied, e.g., transformation-based join~\cite{transform-join-2, warren2006multi, zhu2017auto}, fuzzy-join~\cite{li2021auto, fuzzy-join-1, fuzzy-join-2}, search-join~\cite{lehmberg2015mannheim}, semantic lookup-join~\cite{he2015sema}, etc., which is an interesting area of future work in the context of BI. 

\section{Problem Statement}
\label{sec:problem}
In this section, we first describe preliminaries and the real BI models we harvest, before introducing the \abi problem.

\subsection{Preliminary: BI models}

Business Intelligence is closely related to topics like data warehousing and decision support system, and has been extensively studied. We provide a brief preliminary here and refer readers to surveys like~\cite{kimball2011data, negash2008business, chaudhuri2011overview} for more information. 

\textbf{Fact and dimension tables.}
In data warehousing and BI terminologies, tables involved in BI modeling can be categorized as two types: \textit{fact tables} and \textit{dimension tables}~\cite{kimball2011data}. A \textit{fact table} contains key metrics and measurements of business processes that one intends to analyze (e.g., the revenue of sales transactions). In addition, a fact table contains \textit{foreign keys} that can reference multiple dimension tables, where each \textit{dimension table} contains detailed information associated with the measurements from a unique facet (e.g., a ``\code{Product}'' dimension table contains details of products sold, whereas a ``\code{Date}'' dimension table has detailed day/month/year info of transactions, etc.). 
Figure~\ref{fig:schemas} shows a few examples of the BI schemas, with fact and dimension tables in different colors.

Such a separation of fact/dimension tables has many benefits, such as storage/query efficiency, ease of maintenance, etc.~\cite{kimball2011data, negash2008business, chaudhuri2011overview}. The fact/dimension design is a de-facto standard in BI modeling.

\textbf{Star, snowflake, and constellation schemas.} In BI modeling, fact/dimension tables are frequently organized into what is known as \textit{star/snowflake/constellation schemas}~\cite{kimball2011data}, like shown in Figure~\ref{fig:schemas}.

\textit{Star-schema} refers to the cases where there is one fact table, whose foreign-key columns refer to primary-keys from one or more (non-hierarchical) dimension tables, as illustrated by Figure~\ref{fig:schemas}(a). 

\textit{Snowflake-schema} generalizes the  star-schema, with dimension tables referring to each other in a  hierarchical manner. For example, in Figure~\ref{fig:schemas}(b), the ``\code{Customer}'' dimension refers to a coarser-grained dimension ``\code{Cust-Segment}''. Similarly an ``\code{Address}'' dimension can refer to a coarser-grained ``\code{City}'', which in turn refers to ``\code{Country}''.

While there is only one fact table in star and snowflake schemas, \textit{constellation-schema} generalizes to the cases with multiple fact tables, as shown in Figure~\ref{fig:schemas}(c). 

We note that these three types of schemas are extensively studied in the literature~\cite{kimball2011data, negash2008business, chaudhuri2011overview} and widely adopted in practice. 

\subsection{Harvest Real BI Models}
\label{subsec:data}
In order to understand real BI models created in the wild, and systematically evaluate the effectiveness of \abi, we harvest over 100K real BI models  from public sources that are created using a popular tool Power BI~\cite{power-bi}, whose model files have a suffix ``\code{.pbix}''.

We crawl these ``\code{.pbix}'' model files from two sources. The first is GitHub, where developers and data analysts upload their Power BI models for sharing and versioning. We crawl a total of 27K such model files from GitHub. As a second source, we go through the URLs crawled by a commercial search engine that can lead to ``\code{.pbix}'' model files. Most of these URLs are not crawled by the search engine so we crawl ourselves and obtain 86K model files. 

Combining data from the two sources gives us a large collection of 100K+ real BI models, which cover diverse BI use cases, including financial reporting, inventory management, among many others. 

These real BI models become a valuable dataset for rigorous evaluation -- specifically, from each ``\code{.pbix}'' model  file, we programmatically extract all tables used in the model, 
as well as the ground-truth BI model (join relationships) manually specified by users. 
This enables us to thoroughly evaluate different algorithms using real BI models in the wild, which turn out to be more challenging than synthetic TPC benchmarks used in prior work that tend to be clean and simple.

\subsection{Problem Statement: \abi}

We now define the high-level \abi problem as follows. 

\begin{definition}
\label{def:abi-basic}
[\abi{}]. Given a set of input tables $\mathbf{T} = \{T_1, T_2,$ $\ldots,$ $T_n\}$ used for BI modeling, where each table $T_i$ consists of a list of columns $T_i = (c_{i1}, c_{i2}, \ldots, c_{im})$. Predict  a desired BI model $M(\mathbf{T})$  that consists of a set of joins $M(\mathbf{T})$ $= \{ J_{ij}, J_{pq}, \ldots \}$, where each join $J_{ij}$ is a join between $T_i$ and $T_j$, in the form of $J_{ij}$ $= \left( (c_{ik}, c_{il}, \ldots), (c_{jr}, c_{js}, \ldots) \right)$.
\end{definition}

Note that the output $M(\mathbf{T})$ is restricted to equi-joins for now, 
which can be single-column joins, 
or multi-column joins in the form of $J_{ij}$ $= \left( (c_{ik}, c_{il}, \ldots), (c_{jr}, c_{js}, \ldots) \right)$. 
Also note that we only state the high-level problem so far, and will defer the exact graph-based formulation to Section~\ref{subsec:abi}.

A more general version of the problem goes beyond equi-joins in Definition~\ref{def:abi-basic} to handle complex forms of joins such as transformation-joins~\cite{zhu2017auto} and fuzzy-joins~\cite{li2021auto}, which we will leave as future work.
\iftoggle{fullversion}
{
}
{
}


\section{\abi}
\label{sec:abi}
We will start with an overview of our \abi architecture, before discussing how we predict joins  holistically on graphs.

\subsection{Architecture Overview}

\begin{figure}[tb]
\vspace{-5mm}
	\centering
	\includegraphics[width=0.9\linewidth]{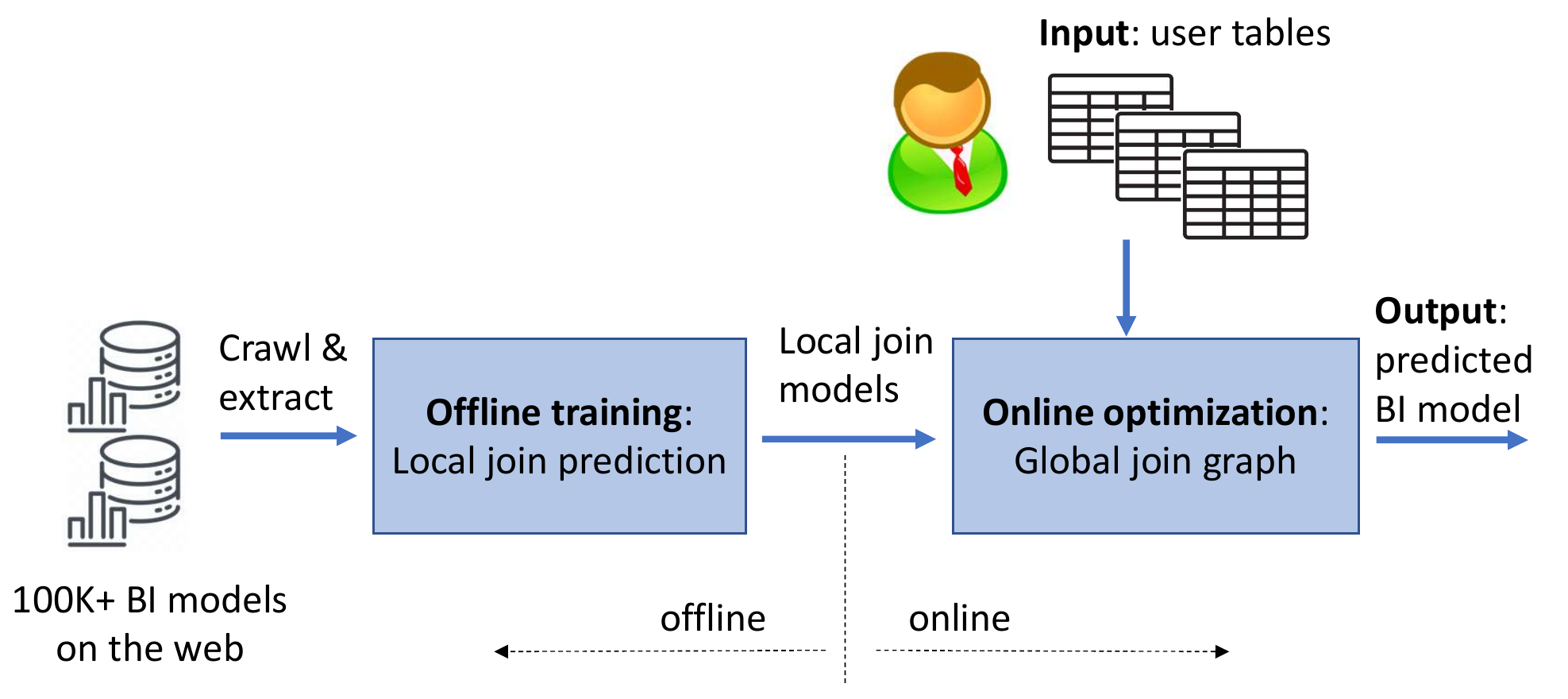}
	\caption{Architecture overview of \abi}
	\label{fig:architecture}
\end{figure}

Figure~\ref{fig:architecture} shows the overall architecture of our \abi system. It consists of an offline component and an online component, as depicted by the two boxes shown in the figure. 

In the offline step, using real BI models we harvested and their ground-truth BI join models, we perform offline training to learn what we call ``local join models'' that can predict, for a given pair of table columns, whether the two are likely joinable.  We use the term ``\textit{local}'' here, because the predictions are pair-wise between two columns, which is in contrast to the ``\textit{global}'' decision at the entire graph level (the focus of this work).

Since the problem of predicting local joinability using data has been studied in other contexts (e.g.,~\cite{fk-ml, yan2020auto}), we do not regard this as our key contribution in \abi, so we will only briefly highlight the important optimizations we make here (e.g., label transitivity, and splitting 1:1 and N:1 joins) in Section~\ref{subsec:local-join}. 

The online step is the key of \abi. Here, given a set of tables (modeled as vertices in a graph), we leverage the local-join prediction models trained offline to ``score'' the joinability of each pair of columns/tables, using calibrated probabilities (which are then modeled as edges in the graph). We  formulate \abi on the resulting graph as a novel graph problem  $k$-MCA, which finds the most probable sub-graph that maximizes the joint probability of all edges, subject to graph-structure constraints. We prove the hardness of the problem and develop efficient algorithms.  We will describe this online part in Section~\ref{subsec:abi}.

\vspace{-1em}
\subsection{Join Prediction: Train Local Classifier}
\label{subsec:local-join}

In this step, given  two lists of columns $C_i \subseteq T$, $C_j \subseteq T'$,
we need to predict the probability that $(C_i, C_j)$ is joinable. (Note that $C_i, C_j$ can be single-columns, but are in general lists of columns for multi-column joins). 
This problem of predicting local joins has been studied in other contexts~\cite{fk-ml, yan2020auto}. We  do not regard this as a new contribution, and will only briefly describe the overall process.

\textbf{The prediction task.} At a high-level, 
given any two candidate columns $(C_i, C_j)$, our task is to predict the corresponding ``joinability'' label, denoted by  $L_{ij}$, where $L_{ij} = 1$ if $(C_i, C_j)$ joins, and $L_{ij} = 0$ otherwise. Since we harvested large amounts of real BI models, each of which contains both data tables and ground-truth joins (programmed by human users), we can use this rich collection of data to produce training data of the form $\{(C_i, C_j), L_{ij}$\}, where $L_{ij}$ corresponds to the actual joined vs. not-joined ground-truth between the two column, which can be programmatically extracted from the BI models we harvest.

This naturally leads to a supervised ML formulation, where we featurize $(C_i, C_j)$ both at the schema-level (e.g., column-header similarity using standard string distance functions such as Jaccard and Edit, as well as pre-trained embedding-based similarity such as SentenceBERT~\cite{SentenceTransformers}), and at the content-level (e.g., column-value overlap based on Jaccard and Containment), for a total of 21 features. In the interest of space,
we leave detailed descriptions of these features, as well as two unique optimizations we develop in this work: (1) separate N:1/1:1 joins, and (2) apply label transitivity, to 
\iftoggle{fullversion}
{
Appendix~\ref{apx:local-join-two-opt} and Appendix~\ref{apx:features}.
}
{
our technical report~\cite{full}.
}

\textbf{Calibrate classifier scores into probabilities.}
After we train the feature-based model using data extracted from real BI models, given a new pair of columns $(C_i, C_j)$, we can use the model to produce classifier-scores and predict the joinability of $(C_i, C_j)$. However, the scores so produced are still heuristic in nature -- e.g., a 0.5 classifier score does not necessarily corresponds to a true join-probability of 0.5, or a 50\% chance of the join being correct. 

In order to make a principled global decision at the graph level, we ``calibrate'' the classifier-scores into true probabilities, using calibration techniques from the ML literature~\cite{niculescu2005predicting}. For any column pair $(C_i, C_j)$, the calibration step produces $P(C_i, C_j)$ that corresponds to the probability of the pair being joinable -- e.g., $P(C_i, C_j)=0.5$ really means that a join prediction between the two columns has 50\% chance of being correct. 
As we will see, this gives a precise probabilistic interpretation, which is important when we reason about the most probable global join graph.

\subsection{\abi: Exploit Global Join Graph}
\label{subsec:abi}

\begin{figure*}[tb]
\vspace{-18mm}
	\centering
	\begin{minipage}[t]{0.48\linewidth}
   \centering
	\includegraphics[width=0.53\linewidth]{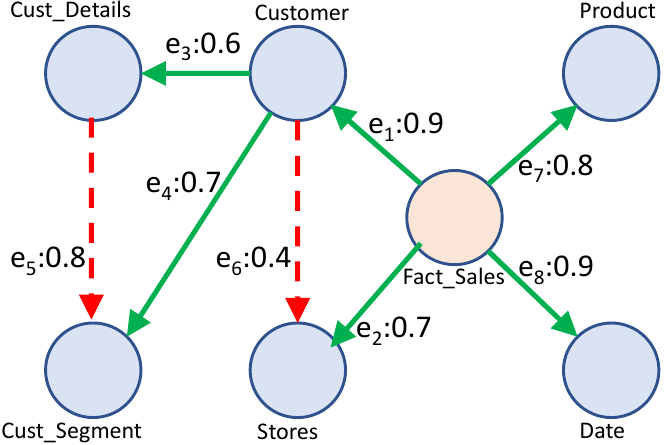}
\vspace{1mm}
\caption{Solve 1-MCA on the graph representation of the tables in Figure~\ref{fig:schemas}(b), with join candidates as edges.}
	\label{fig:join-graph-snowflake}
	\end{minipage}
 \hfill
  \hfill
	\begin{minipage}[t]{0.48\linewidth}
   \centering	\includegraphics[width=0.95\linewidth]{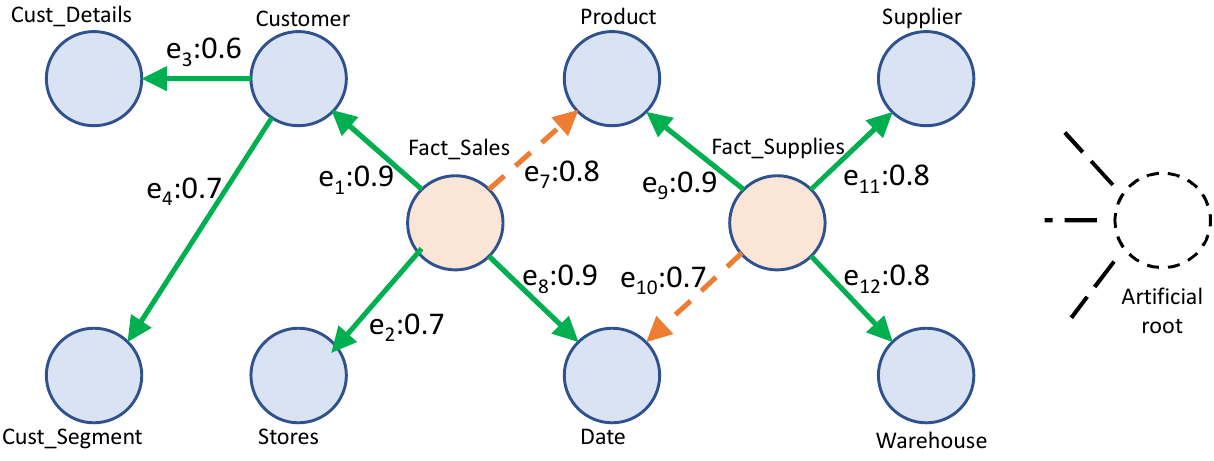}
\vspace{1mm}
	\caption{Solve $k$-MCA on the graph representation of the tables in Figure~\ref{fig:schemas}(c), with join candidates as edges.}
	\label{fig:join-graph-constellation}
	\end{minipage}
 \vspace{-1em}
\end{figure*}

We are now ready to solve the \abi problem. 
We first introduce how we represent tables and candidate joins on a global join graph, before describing our graph formulation.
\subsubsection{\textbf{Representing relationships in a global graph.}} \hfill\\
Given a set of tables $\mathbf{T}$ and possible joins, we can construct a directed graph $G=(V, E)$ as follows. We represent each input table $T \in \mathbf{T}$ as a vertex $v(T) \in V$, 
and each possible join candidate between columns $(C_i, C_j)$ as a weighted  edge $e_{ij} \in E$, where the edge weight $w(e_{ij})$ is simply the calibrated join probability $P(C_i, C_j)$, produced by our local classifier (Section~\ref{subsec:local-join}), which scores every pair of candidate columns whose containment is over a threshold. We follow the convention to use a directed edge $e_{ij}$ to represent N:1 joins, which point from N-side (FK) columns $C_i$, to the 1-side (PK) columns $C_j$. We represent 1:1 joins as bi-directional edges.

\begin{example}
\label{ex:graph}
Figure~\ref{fig:join-graph-snowflake} shows a graph representation of the  tables in Figure~\ref{fig:schemas}(b), where each vertex corresponds to a table. We mark the ground-truth joins in Figure~\ref{fig:schemas}(b) as solid green edges in Figure~\ref{fig:join-graph-snowflake}, while other candidate joins not in ground-truth as dotted red edges. 

For instance, the dotted edge $(e_5: 0.8)$ represents an candidate join between the column ``\code{Customer-ID}'' (in table ``\code{Cust-Details}''), and column ``\code{Customer-Segment-ID}'' (in table ``\code{Cust-Segments}'').
Note that the column pair (``\code{Customer-ID}'', ``\code{Customer-Segment-ID}'') should not join because they refer to two semantically different types of IDs, which however may appear like a plausible join to Local-Classifier (because of high name-similarity and value-overlap), which leads to a high Local-Classifier score (0.8). A greedy method that focuses on promising edges locally can incorrectly predict this false-positive join, which is a mistake that global methods like \abi can prevent.
\end{example}

\subsubsection{\textbf{Precision Mode (k-MCA-CC)}} \hfill\\
\label{subsec:abi-p}
We are now introducing our formulation using the graph. At a high level, we operate in two steps: (1) a ``precision-mode'' stage where we focus on finding  the salient snowflake-like structures that are the ``backbones'' of the underlying schema graph (which ensures high precision thanks to the graph-structure constraints it imposes); 
and (2) a ``recall-mode'' stage that complements the precision-mode, by finding additional joins beyond typical snowflakes. We will  introduce both in turn below.

We first introduce our precision-mode formulation, referred to as $k$-MCA-CC. For ease of exposition, we will illustrate the thought process of arriving at $k$-MCA-CC in 3 steps: (1) We will start with a simplistic assumption that the schema graph has exactly one snowflake, which we model with a graph-theoretical problem called 1-MCA. (2) We then generalize this to arbitrary numbers of snowflakes, using a formulation we call $k$-MCA. (3) Finally, we generalize $k$-MCA to include additional graph-structure constraints to arrive at our final formulation, which we call $k$-MCA-CC. We summarize our key results in this section in Table~\ref{tab:mca-result-summary} for reference.

\begin{table}[t]
\label{tab:mca-result-summary}
\footnotesize
    \centering
    \begin{tabular}{|c|>{\centering\arraybackslash}m{2.2cm}|>{\centering\arraybackslash}m{2cm}|>{\centering\arraybackslash}m{2cm}|}
    \hline
    & known results & \multicolumn{2}{c|}{new results}   \\ \hline
    Problem & \textbf{1-MCA} & \textbf{k-MCA} & \textbf{k-MCA-CC}   \\ \hline
    Description & find the most probable 1-snowflake & find the most probable k-snowflakes & most probable k-snowflakes w/ constraints \\ \hline
    Hardness & Poly-time solvable~\cite{chu1965shortest} & Poly-time solvable (Theorem~\ref{thm:k-mca-optimal})  & EXP-APX-hard (Theorem~\ref{thm:kmcacc-inapprox}) \\ \hline
    Algorithm & Chu-Liu/Edmond's algorithm~\cite{chu1965shortest} &  \shortstack[c]{ours \\ (Algorithm~\ref{alg:k-mca})}  & \shortstack[c]{ours \\ (Algorithm~\ref{alg:k-mca-cc})}  \\ \hline
    \end{tabular}
 \caption{Summary of results for MCA problem variants.} 
    \label{tab:mca-result-summary}
\end{table}

\textbf{(1) Exactly one snowflake: 1-MCA}.
To start with, we discuss the simplistic case where we know there is exactly one snowflake in the underlying graph (e.g., Figure~\ref{fig:schemas}(b)). Note that this is only for ease of illustration, and not an actual assumption that we rely on (we will show how this can be relaxed next). 

Given a graph $G=(V, E)$ where candidate joins are marked as edges like in Figure~\ref{fig:join-graph-snowflake}. Since we know there is one snowflake schema, intuitively we want to select edges (joins) $J \subseteq E$, such that: 

(a) The graph induced by $J$,  $G'=(V, J)$, is a snowflake that connects all vertices in $V$; 

(b) If more than one such snowflake-structure exists, find the most probable snowflake, based on the joint-probability of all joins selected in $J$, $P(J) =$ $\prod_{e_{ij} \in J} P(C_i, C_j)$. 

These two considerations can be written as the following optimization problem, which we refer to as 1-Most-Probable-Snowflake (1-MPS):
\begin{align}
\text{(1-MPS)} \qquad{} \max_{J \subseteq E} &~~ \prod_{e_{ij} \in J} P(C_i, C_j) \label{eqn:1mps_obj} \\  
\mbox{s.t.} ~~ & G'=(V, J) ~~ \text{is a snowflake}  \label{eqn:1mps_snowflake}
\end{align}
Note that the constraint in  Equation~\eqref{eqn:1mps_snowflake}  of 1-MPS corresponds to requirement (a), while the objective function in Equation~\eqref{eqn:1mps_obj} maps to requirement (b).

Since snowflake is used informally  in  BI, to make it more formal we map it to a structure in graph theory called \textit{arboresence}~\cite{fournier2013graphs}. 

\begin{definition}
\label{def:arborescence} [Arborescence].
A directed graph $G=(V, E)$ is called an  \textit{arborescence}, if there is a unique vertex $r \in V$ known as the root, such that there is exactly one directed path from $r$ to every other $v \in V, v\neq r$. Equivalently, a directed graph $G$ is an arborescence if all its vertices have in-degree of 1, except a unique root vertex $r \in V$ that has in-degree of 0~\cite{fournier2013graphs}.
\end{definition}

Intuitively, arborescence is a directed rooted tree where all its vertices are pointing away from the root. We use an example to show the relationship between snowflakes and arborescence.

\begin{example} 
Consider the sub-graph $G$ induced by all green edges in Figure~\ref{fig:join-graph-snowflake}, where each green edge would correspond to a ground-truth join in the snowflake schema of Figure~\ref{fig:schemas}(b). This sub-graph $G$ is an arborescence, because if we take the vertex marked as ``\code{Fact\_Sales}'' as the root $r$, then there is exactly one directed path from $r$ to every other vertex in $G$.  Equivalently, we can check that this root $r$ has in-degree of 0, while all other vertices in $G$ have in-degree of exactly 1, which also ensures that $G$ is an arborescence.
\end{example}

Recall that we know there is exactly one snowflake in 1-MPS, which is equivalent to saying that the sub-graph induced by $J \subseteq E$ is an arborescence. 
We rewrite 1-MPS into 1-MPA (1-Most-Probable-Arborescence), defined as follows.
\begin{align}
\text{(1-MPA)} \qquad{} \max_{J \subseteq E} &~~ \prod_{e_{ij} \in J} P(C_i, C_j) \label{eqn:1mpa_obj} \\  
\mbox{s.t.} ~~ & G'=(V, J) ~~ \text{is an arborescence}  \label{eqn:1mpa_snowflake}
\end{align}

\begin{example}
\label{ex:1-mpa}
We revisit the graph in Figure~\ref{fig:join-graph-snowflake}. Using the formulation of 1-MPA, the best arborescence with the highest joint probability (correspondingly, the most probable snowflake), is the set of solid edges marked in green, denoted by $J^* = \{e_1, e_2, e_3, e_4, e_7, e_8\}$, whose joint probability is $0.9 * 0.7 * 0.6 * 0.7 * 0.8 * 0.9$. It can be verified that $J^*$ is the most probable arborescence among all arborescences (that span the entire vertex set), which corresponds to the ground-truth snowflake schema shown in Figure~\ref{fig:schemas}(b). 

Consider an alternative arborescence, such as $J' = \{e_1, e_3,$ $e_5, e_6,$ $e_7, e_8\}$, which removes $e_2, e_4$ from $J^*$ and adds $e_5, e_6$ (marked in dotted red lines). Observe that this new $J'$ also forms a 1-arborescence based on Definition~\ref{def:arborescence}. However, $J'$ has a lower joint probability than $J^*$  (because $e_5, e_6$ in $J'$ has a joint probability of $0.8 * 0.4$=0.32, while $e_2, e_4$ in $J^*$ has a joint probability of $0.7 * 0.7$=0.49), which will thus not be selected based on the 1-MPA formulation above. 

We note that in 1-MPA, $J^*$ is selected based on a global decision of optimality, which avoids locally-greedy decisions -- e.g., the incorrect $e_5$ will not be selected despite its high score like we mentioned in Example~\ref{ex:graph}.
\end{example}

While we now use arborescence to formalize 1-MPA, it still has a cross-product that is not amenable to optimization. We thus perform the following transformation: instead of assigning join-probability $P(C_i, C_j)$ as the edge weight for each edge $e_{ij}$, we perform a logarithmic transformation and set the edge weight of each $e_{ij}$ as:
\begin{equation}
\label{eqn:edge-log}
w(e_{ij}) = -\log(P(C_i, C_j))
\end{equation}
Using the new  transformed $w(e_{ij})$, for each instance of the 1-MPA problem, we construct a new minimization problem below that we term 1-MCA, which uses summation in the objective function:
\begin{align}
\text{(1-MCA)} \qquad{} \min_{J \subseteq E} &~~ \sum_{e_{ij} \in J} w(e_{ij}) \label{eqn:1mca_obj} \\  
\mbox{s.t.} ~~ & G'=(V, J) ~~ \text{is arboresence} \label{eqn:1mca_arborescence}
\end{align}
It can be shown that solving 1-MPA is equivalent to solving the corresponding 1-MCA. 

\begin{lemma}
\label{lem:eq}
A solution $J^* \subseteq E$ is an optimal solution to 1-MPA, if and only if $J^*$ is an optimal solution to the corresponding 1-MCA. 
\end{lemma}

\begin{proof}
First, observe that 1-MPA and 1-MCA have identical feasible regions, as they are subject to the same constraints.

Next, we show that an optimal solution $J^*$ that maximizes the objective function $\prod_{e_{ij} \in J^*} P(C_i, C_j)$ in Equation~\eqref{eqn:1mpa_obj} of 1-MPA, will simultaneously minimize the objective function $\sum_{e_{ij} \in J} w(e_{ij})$ in Equation~\eqref{eqn:1mca_obj} of 1-MCA. This is the case because the objective function of 1-MCA is:~~ $\sum_{e_{ij} \in J^*} w(e_{ij}) $ $ = $ $\sum_{e_{ij} \in J^*} -\log\left(P(C_i, C_j)\right)$ $= -\log(\prod_{e_{ij} \in J^*} P(C_i, C_j))$, where the term inside $-\log()$ is exactly the objective function of 1-MPA, thus ensuring that  
1-MCA is minimized if and only if 1-MPA is maximized, and completes the proof.
\end{proof}

The reason we construct 1-MCA for each instance of 1-MPA, is that 1-MCA relates to a known problem in graph theory called ``\textit{Minimum-Cost-Arborescence}'' (abbreviated as MCA\footnote{Although MCA is not a well-known concept, we note that MCA is directly analogous to a better-known problem in graph theory called ``\textit{Minimum-Spanning-Tree} (MST)''~\cite{karger1995randomized}, with the only difference that MST is defined on undirected graphs whereas MCA is on directed graphs.})~\cite{edmonds1967optimum}, which finds a spanning arborescence (covering all vertices) in a directed graph $G$ that has the smallest edge-weights. Note that in the simplistic setting where we know there is exactly 1 snowflake (arborescence), our 1-MCA problem directly translates to the MCA problem~\cite{edmonds1967optimum}. Since MCA is known in graph theory with polynomial-time solutions called the Chu–Liu/Edmonds' algorithm~\cite{edmonds1967optimum, chu1965shortest}, 
our construction 
allows us to solve 1-MPA efficiently  by leveraging the same algorithms.

\begin{small}
\begin{algorithm}[t]
\SetKw{kwReturn}{return}
 \Input{all input tables $\mathbf{T}$ in a BI-model}
 \Output{Graph $G=(V, E)$ that represents $\mathbf{T}$}
 
 $V \leftarrow \{v_T | T \in \mathbf{T} \}$, with $v_T$ representing each $T \in \mathbf{T}$ 

 $E \leftarrow \{\}$
 
  \ForEach{$(C_i, C_j)$ satisfying Inclusion-Dependency in $\mathbf{T}$ \label{line:1mca-IND}} 
    {
       $P(C_i, C_j) \leftarrow \text{Local-Classifier}(C_i, C_j)$  \label{line:1mca-LC} 
       
       $w(e_{ij}) \leftarrow -\log(P(C_i, C_j))$ \label{line:construct-graph-edge-weight}

       $E \leftarrow E \cup \{e_{ij}\}$, with edge-weight $w(e_{ij})$
    }

\kwReturn $G(V, E)$
\caption{Construct graph with edge-weights}
\label{alg:build-graph}
\end{algorithm}
\end{small}

We use our running example to show the connection between 1-MPA and 1-MCA.

\begin{example}
\label{ex:1-mca}
Continue with Example~\ref{ex:1-mpa}. To solve the 1-MPA problem for the graph in Figure~\ref{fig:join-graph-snowflake}, 
we use the transformation in Equation~\eqref{eqn:edge-log} and  construct an instance of 1-MCA on the same graph, where all edge-weights are now $w(e_{ij}) = -\log(P(C_i, C_j))$. It can be verified that $J^* = \{e_1, e_2, e_3, e_4, e_7, e_8\}$ is the minimizer of Equation~\eqref{eqn:1mca_obj} in 1-MCA, with the smallest objective-value $-(\log(0.9) + \log(0.7) + \log(0.6) + \log(0.7) + \log(0.8) + \log(0.9))$,   which can be efficiently solved using the Chu–Liu/Edmonds algorithm. Note that $J^*$ is the same optimal solution to 1-MPA, as we see in Example~\ref{ex:1-mpa}.
\end{example}

\underline{Algorithm for 1-MCA.} Given a set of input tables $\mathbf{T}$ (for which a BI-model needs to be built),
the complete pseudo-code to construct a graph $G$ for 1-MCA 
is shown in  Algorithm~\ref{alg:build-graph}. Since we are given tables $\mathbf{T}$, in Line~\ref{line:1mca-IND} we enumerate column-pairs $(C_i, C_j)$  in $\mathbf{T}$ for which Inclusion-Dependencies  (IND)~\cite{casanova1982inclusion} hold approximately, which are possible joins that we should consider (note that efficient IND-enumeration is a standard step~\cite{de2002efficient}, so we invoke existing techniques here). In Line~\ref{line:1mca-LC}, we ``score'' each $(C_i, C_j)$ using our Local-classifier to obtain calibrated probabilities $P(C_i, C_j)$ (Section~\ref{subsec:local-join}), which are transformed in Line~\ref{line:construct-graph-edge-weight} to become edge-weights  $w(e_{ij})$ (Equation~\eqref{eqn:edge-log}). 

Using the resulting graph $G$ constructed from $\mathbf{T}$, we can invoke the Chu–Liu/Edmonds' algorithm, which yields the optimal solution $J^*$ to 1-MCA (and thus also 1-MPA).


We summarize our main result for 1-MCA in the first column of Table~\ref{tab:mca-result-summary}. We note that although we leverage known results from the graph theory here, to the best of our knowledge, we are the first to connect the popular concept of snowflakes in BI, with arborescence in graph theory.

In the following, we will extend 1-MCA to general cases, and develop new results not known in graph theory or databases.

\textbf{(2) Arbitrary number of snowflakes: $k$-MCA}.
So far we use 1-MCA to solve the simplistic case where we know there is exactly 1 snowflake. In general, there can be multiple snowflakes, and we do not know its exact number beforehand (e.g., the ground-truth schema of Figure~\ref{fig:schemas}(c) has 2 snowflakes, which is unknown a priori).

In this section, we extend 1-MCA to the more general $k$-MCA, where the proposed $k$-MCA formulation will not only find the most probable $k$ snowflakes like in 1-MCA, but also infer the right number of snowflakes $k$ at the same time.

We first extend the concept of arborescence in graph-theory (Definition~\ref{def:arborescence}), to the more general \textit{$k$-arboresence} below.

\begin{definition}
\label{def:k-arboresence}
[$k$-arboresence].  A directed graph $G=(V, E)$ is an  $k$-\textit{arborescence}, if its underlying undirected graph has a total of $k$ joint connected-components, written as $\{G_i=(V_i, E_i) | i \in [k]\}$, such that $\bigcup_{i \in [k]}{V_i} = V$, $\bigcup_{i \in [k]}{E_i} = E$. Furthermore, each $G_i$ is an arborescence, for all $i \in [k]$.
\end{definition}

Note that when $k=1$, the notion of 1-arborescence degenerates into arborescence described in Definition~\ref{def:arborescence}. We will henceforth use 1-arborescence and arborescence interchangeably when the context is clear. 

\begin{example}
\label{ex:k-arboresence} 
Consider the graph in Figure~\ref{fig:join-graph-constellation}, which is the graph representation of the constellation (multi-snowflake) schema in Figure~\ref{fig:schemas}(c). The sub-graph with green solid edges (ignoring the dotted edges for now), is a 2-arboresence. It is a  2-arboresence because both of its two connected-components are arboresences (Definition~\ref{def:arborescence}), rooted at ``\code{Fact\_Sales}'' and ``\code{Fact\_Supplies}'', respectively.
\end{example}

Intuitively, we use the $k$-arboresence structure to force the $k$ underlying snowflakes to emerge, which reveals the most salient ``backbones'' of the underlying schema. It should be noted that  in the precision-mode of \abi (this section), because we focus exclusively on finding snowflake/arborescence structures to ensure high precision, some desired joins may be missing in the $k$-arboresence (e.g., the orange dotted edges in Figure~\ref{fig:join-graph-constellation}), which is something that we will get to in the recall-mode of \abi next (Section~\ref{subsec:abi-r}). 

Given that we want to use $k$-arboresence to let the $k$ snowflakes emerge, the next question is how to find the right $k$. 
Conceptually, we can iterate over all possible $k$, and pick the ``best'' $k$, using a suitable ``goodness'' function  (e.g., revealing the right number of snowflakes). 

First, recall that a $k$-arborescence on graph $G=(V, E)$ has exactly $k$ connected components, and is bound to have exactly $(|V|-k)$ edges (because  every vertex except the root has in-degree of 1, per Definition~\ref{def:arborescence}). 

Given these, we can see that while the ``sum of edge-weights'' objective function in Equation~\eqref{eqn:1mca_obj} for 1-MCA  is a suitable ``goodness'' function  to compare two 1-arboresences (both with $k=1$), it is no longer suitable to compare a $k_1$-arboresence and a $k_2$-arboresence (with $k_1 \neq k_2$), because the two will have different number of edges, making the comparison using the sum of edge-weights in Equation~\eqref{eqn:1mca_obj} ``unfair''. In fact, using Equation~\eqref{eqn:1mca_obj} and a flexible $k$ is guaranteed to lead to $|V|$ disconnected vertices (a trivial $|V|$-arborescence) as the optimal, because it has no edges and thus no edge-weights.



To make the comparisons ``fair'' between two $k$-arboresences with different values of $k$, and prevent disconnected components from having artificial advantages, we modify the objective function as follows. 
Since a $k$-arboresences has exactly $k$ connected components and $(|V|-k)$ edges, we imagine that there are $(k-1)$ ``virtual-edges'', each with a parameterized edge-weight $p$, that connect the $k$ connected components into one, such that a $k$-arboresences always has the same number of edges as 1-arboresences, regardless of $k$ (because  $(|V|-k) + (k-1) = (|V|-1)$).   
Accounting for the $(k-1)$ new virtual edges in Equation~\eqref{eqn:1mca_obj} leads to Equation~\eqref{eqn:kmca_obj} below, which is new objective function we use for $k$-MCA:
\begin{align}
\text{($k$-MCA)} \quad{} \min_{\substack{J \subseteq E,~ k \leq |V|}} &~~ \sum_{e_{ij} \in J} w(e_{ij}) + (k-1) \cdot p \label{eqn:kmca_obj} \\  
\mbox{s.t.} ~~ & G'=(V, J) ~~ \text{is a  $k$-arboresence} \label{eqn:kmca_arborescence}
\end{align}
The parameter $p$ we introduce effectively controls the number of snowflakes (e.g., a larger $p$ would
``penalize'' having more disconnected snowflakes). The question is how to set $p$ appropriately. 
Recall that we use calibrated join-probability $P(C_i, C_j)$ to produce edge weight $w(e_{ij}) = -\log(P(C_i, C_j))$, then naturally the virtual edges should be imagined as edges with a join-probability of exactly $0.5$, which means a 50\% chance of being joinable and thus a coin-toss that is ok to either include or exclude (which makes them ``virtual'').  

As another way to look at this, consider  when we drop a regular edge $e_{ij}$ from  a $k$-arboresence to create a $(k+1)$-arboresence -- the objective function of the latter in Equation~\eqref{eqn:kmca_obj} will have the edge-weight $w(e_{ij})$ removed, but incur a penalty cost of $p$ from the additional virtual-edge we introduce. When the penalty-weight $p$ on the virtual-edge corresponds to a join-probability of 0.5, it would discourage $e_{ij}$ from being dropped if its $P(C_i, C_j)$ is over 0.5 (50\% chance of being joinable), and encourage $e_{ij}$ to be dropped if its $P(C_i, C_j)$ is under 0.5, which is exactly the right thing to do.

We, therefore, use $p = -\log(0.5)$ as our natural choice of penalty weight in $k$-MCA. Our experiments suggest that empirically $0.5$ is indeed the right choice (Section~\ref{sec:exp}), thanks to the fact that we use true calibrated probabilities (Section~\ref{subsec:local-join}).

Observe that if we know $k=1$ a priori, our $k$-MCA degenerates exactly into 1-MCA like before. When $k$ is unknown however, the objective function in Equation~\eqref{eqn:kmca_obj} would help reveal the best $k$-snowflakes, as well as the right number of $k$. We show this intuitively using the example below.


\begin{example}
\label{ex:kmca}
We revisit Figure~\ref{fig:join-graph-constellation} where all edges are possible join candidates, and solve $k$-MCA on this example graph. 

First, observe that there is no 1-arboresence in this graph. Let $J^*$ be all the green edges, then the subgraph induced by $J^*$ is a 2-arboresence (rooted at the two fact-tables). Since $k=2$, the penalty term in this case is $(2-1) (-\log(0.5)) = 1$. It can be verified that   $J^*$ has the lowest cost in all  2-arboresences.

It is possible to have 3-arboresences here too -- for example if we remove $e_1$ from $J^*$, then the remaining green-edges in $J' = J^* \setminus \{e_1\}$ induce a 3-arboresence. It can be verified that the cost of $J'$ is higher than that of $J^*$,  because $J'$  removes $e_1$ from $J^*$ and thus lowers its cost by $w(e_1) = -\log(0.9) = 0.13$, but incurs a higher penalty-cost of $(3-1) (-\log(0.5)) = 2$,  which makes $J'$ less desirable than $J^*$.
\end{example} 

\underline{Algorithm for $k$-MCA.} A naive way to solve $k$-MCA that builds on top of 1-MCA above, is to enumerate different values of $k$, and for each $k$, exhaustively enumerate  all $k$-way partitions of $G$, then invoke 1-MCA on each resulting graph partition, to find the optimal $k$-MCA. This approach is straightforward but inefficient (because $k$ can be as large as $|V|$, making it exponential in $|V|$).

We design an algorithm that solves $k$-MCA, using a graph-based construction that reduces any instance of the new $k$-MCA problem into \textit{one} instance of $1$-MCA (which admits efficient solutions~\cite{chu1965shortest}). Specifically, given a graph $G=(V, E)$ on which $k$-MCA needs to be solved, we introduce a new vertex $r$ that is an ``artificial root'', and connects $r$ with all $v \in V$ using edges $e(r, v)$, with edge-weight $w(e(r, v))=p$. This leads to a new  constructed graph $G'=(V', E')$ where $V' = V \cup \{r\}$,  $E' = E \cup \{e(r, v)| v \in V\}$. 

We solve $1$-MCA  on the new $G'$, and let $J^*_1$ be the optimal 1-MCA. Let $J^*_k = J^*_1 \setminus \{e(r, v) | v \in V\}$ be edges in $J^*_1$ not incident with the artificial-root $r$. We can show that $J^*_k$ is the optimal solution to $k$-MCA on the original $G$. 
The complete steps of this algorithm are shown in Algorithm~\ref{alg:k-mca}.

\begin{small}
\begin{algorithm}[t]
\SetKw{kwReturn}{return}
 \Input{Graph $G=(V, E)$}
 \Output{optimal k-MCA (k-snowflakes)}
 
 
       


$V' \leftarrow V \cup \{r\}$

$E' \leftarrow E \cup \{e(r, v)| v \in V\}$, \text{with} $w(e(r, v)) = p$

$J^*_1 \leftarrow$ solve 1-MCA on $G'=(V', E')$ with Chu-Liu/Edmonds' algo \label{line:kmca-solve-1mca} 

$J^*_k = J^*_1 \setminus \{e(r, v) | v \in V\}$ \label{line:kmca-return}

\kwReturn $J^*_k$
\caption{Solve $k$-MCA for constellation schema}
\label{alg:k-mca}
\end{algorithm}
\end{small}

\begin{theorem}
\label{thm:k-mca-optimal}
Algorithm~\ref{alg:k-mca} solves $k$-MCA  optimally, in time polynomial to the input size.
\end{theorem}

\begin{proof}
Given a graph $G=(V, E)$, let $J^*_k$ be the optimal $k$-MCA of $G$, with objective function value $c_k(J^*_k)$, where $c_k$ is the objective function defined in Equation~\eqref{eqn:kmca_obj}. Because $J^*_k$  is a $k$-arboresence, by Definition~\ref{def:k-arboresence}, $J^*_k$ has exactly $k$ disjoint components $\{G_1, G_2, \ldots G_k\}$, all of which are 1-arboresence, with $R = \{r_1, r_2, \ldots r_k\}$ as roots.

We show that on our constructed graph $G' = (V', E')$, the solution $J^*_1 = J^*_k \cup \{(r, r_i) | r_i \in [k]\}$ must be an optimal solution to the 1-MCA problem on $G'$.

We show this by contradiction. Suppose $J^*_1$ is not an optimal solution to $G'$, which means that there is another $J'_1$ with a smaller cost as defined in the 1-MCA objective function. Let the objective function in Equation~\eqref{eqn:1mca_obj} be $c_1$, we can write this as 
\begin{equation}
\label{eqn:proof_expand_0}
c_1(J'_1) < c_1(J^*_1)
\end{equation}
In the solution $J'_1$, let $R'$ be the set of vertices incident with artificial root $r$. Removing from $J'_1$ the set of edges incident with $r$ produces $J'_k = J'_1 \setminus \{e(r, r') | r' \in R'\}$.  $J'_k$ is an $|R'|$-arboresence, and thus a feasible solution to $k$-MCA on $G$. Similarly, removing from $J^*_1$ the set of edges incident with $r$ produces $J^*_k = J^*_1 \setminus \{e(r, r') | r' \in R\}$, which is also a feasible solution to $k$-MCA on $G$.

Since $J^*_k$ is an $|R|$-arboresence and $J'_{k}$ is an $|R'|$-arboresence, by the definition of $c_k$ in Equation~\eqref{eqn:kmca_obj}, we know 
\begin{equation}
\label{eqn:obj_expand_1}
c_k(J^*_k) = \sum_{e\in J^*_k}{w(e)} + (|R|-1)p
\end{equation}
\begin{equation}
\label{eqn:obj_expand_2}
c_k(J'_k) = \sum_{e\in J'_k}{w(e)} + (|R'|-1)p
\end{equation}
Because we know $J^*_k$ is an optimal solution to $k$-MCA, we know $c_k(J^*_k) \leq c_k(J'_k)$. Combining with Equation~\eqref{eqn:obj_expand_1} and~\eqref{eqn:obj_expand_2}, we get:  
\begin{equation}
\label{eqn:obj_expand_3}
\sum_{e\in J^*_k}{w(e)} + (|R|-1)p \leq \sum_{e\in J'_k}{w(e)} + (|R'|-1)p
\end{equation}
Adding $p$ to both sides of Equation~\eqref{eqn:obj_expand_3}, we get $\sum_{e\in J^*_k}{w(e)} + (|R|)p \leq \sum_{e\in J'_k}{w(e)} + (|R'|)p$. Note that the left-hand-side is exactly the objective-value of $J^*_1$ on $G'$, or $c_1(J^*_1)$, while the right-hand-side is the  objective-value of $J'_1$ on $G'$. This gives $c_1(J^*_1) \leq c_1(J'_1)$, contradicting with  our assumption in Equation~\eqref{eqn:proof_expand_0}, thus proving that  $J^*_1$ must be an optimal solution to 1-MCA on $G'$.

Since $J^*_1$ can be solved optimally in Line~\ref{line:kmca-solve-1mca} of Algorithm~\ref{alg:k-mca} for 1-MCA in polynomial time (Line~\ref{line:kmca-solve-1mca}), and all other steps in Algorithm~\ref{alg:k-mca} also take time polynomial in the input size, we conclude that Algorithm~\ref{alg:k-mca}  can solve $k$-MCA optimally, in polynomial time.
\end{proof}

\begin{example}
\label{ex:kmca_algo}
We continue with Example~\ref{ex:kmca}. Using Algorithm~\ref{alg:k-mca}, we would construct a new graph $G'$ by adding an artificial root $r$ (shown on the right of Figure~\ref{fig:join-graph-constellation}), which connects to all existing vertices $v_i$ with an edge $e(r, v_i)$ with the same penalty weight $w(e) = p$. Using Line~\ref{line:kmca-return} of Algorithm~\ref{alg:k-mca} to solve the 1-MCA on $G'$ produces the optimal solution of $J_1^*$ that consists of all solid green edges, plus two artificial-edges connecting $r$ with the two fact-table, which can be verified is the optimal 1-MCA.  Line~\ref{line:kmca-return} of Algorithm~\ref{alg:k-mca} would  then produce $J_k^*$ corresponding to of all green edges (removing the two artificial-edges connecting fact-tables to the artificial-root). It can be verified that $J_k^*$ in this example  is the optimal $k$-MCA, which is also the ground-truth schema in Figure~\ref{fig:schemas}(c).
\end{example}

\textbf{(3) Arbitrary number of snowflakes with constraints: k-MCA-CC}. The $k$-MCA formulation we considered so far can handle arbitrary numbers of snowflakes. This however, can be further improved by adding an additional constraint on the graph that we call ``FK-once''.\footnote{Other possible constraints that can benefit the inference, such as ``no-cycles'' of join paths in inferred schema graph, is implicit from the definition of arboresence.} 

FK-once refers to the property that the same FK column in a fact-table should likely not refer to two different PK columns in two dimension tables (a common property known in the database literature~\cite{chen2014fast, hpi}). On a graph, such a structure would correspond to two edges $(C_i, C_j), (C_i, C_m)$, pointing from the same $C_i$, to two separate $C_j$ and $C_m$, which usually indicates that one of the joins is incorrect. 
For example,  the same FK column ``\code{Customer-ID}'' in  ``\code{Sales}'' table may appear joinable with both (1) the PK column ``\code{C-ID}'' of the ``\code{Customers}'' table, and (2) the PK ``\code{Customer-Segment-ID}'' of the ``\code{Customer-Segments}'' table (because both have high column-header similarity and value-overlap). However, we know that an FK should likely only join one PK (or otherwise we have two redundant dimension tables).

We add the FK-once constraint as a cardinality-constraint (CC) to $k$-MCA, which leads to $k$-MCA-CC below:
\begin{align}
\hspace{-4mm}
 \text{(k-MCA-CC)} 
\min_{J \subseteq E,~k \leq |V|} &~~ \sum_{e_{ij} \in J} w(e_{ij}) + (k-1) \cdot p \label{eqn:kmcacc_obj} \\  
\mbox{s.t.} ~~ & G'=(V, J) ~~ \text{is  $k$-arboresence} \label{eqn:kmcacc_kmca}  \\
~~ &  i \neq l, \forall e_{ij} \in J, e_{lm} \in J, j \neq m \label{eqn:kmcacc_fk_once} 
\end{align}

Note that Equation~\eqref{eqn:kmcacc_fk_once} is the new FK-once constraint, which states that no two edges $e_{ij}$, $e_{lm}$ in the selected edge-set $J$ should share the same starting column-index, or $(i\neq l)$.\footnote{Note that the constraint is on column-index and at a column-level -- one table can still have multiple different FK columns pointing from the same \textit{table/vertex}, to different PK columns of different \textit{tables/vertices}.}

\begin{theorem}
\label{thm:kmcacc-inapprox}
$k$-MCA-CC is NP-hard. Furthermore, it is Exp-APX-complete, making it inapproximable in polynomial time.
\end{theorem}

\iftoggle{fullversion}
{
We prove Theorem~\ref{thm:kmcacc-inapprox} using a reduction from non-metric min-TSP~\cite{escoffier2006completeness}. Proof of this can be found in 
Appendix~\ref{apx:proof-kmcacc-inapprox}.
}
{
We prove Theorem~\ref{thm:kmcacc-inapprox} using a reduction from non-metric min-TSP~\cite{escoffier2006completeness}. Proof of this can be found in 
our technical report~\cite{full}.
}
It is interesting to see that adding one constraint in $k$-MCA-CC makes it considerably more difficult than $k$-MCA (which however is important in terms of result quality, as we will show in experiments).

\underline{Algorithm for $k$-MCA-CC.}
Despite the hardness, we develop a novel algorithm that solves $k$-MCA-CC optimally, leveraging the branch-and-bound principle~\cite{branch-and-bound}, and the sparsity of join edges pointing from the same columns. 

Algorithm~\ref{alg:k-mca-cc} shows the pseudo-code of this recursive procedure. We first invoke Algorithm~\ref{alg:k-mca} to solve the unconstrained version $k$-MCA  and obtain $J$ (Line~\ref{line:kmcacc-solve-kmca}). We check $J$ for constraint violations (Line~\ref{line:kmcacc-check}) -- if there is no violation, we are done and $J$ is the optimal solution to $k$-MCA-CC. Alternatively, if $J$ has constraint violations, but the cost of the unconstrained version $c(J)$ is already higher than a best solution to $k$-MCA-CC found so far (Line~\ref{line:kmcacc-check-cost}), we know adding the constraints will only degrade solution quality so we return null for this solution space. Otherwise, in Line~\ref{line:kmcacc-else}, let $C_s = \{ e_{sj}, e_{sk}, \ldots \} \subseteq J$ be one set of conflicting edges in $J$ from the same column 
(thus violating the FK-once constraint). We partition $C_s$ into $|C_s|$ number of subsets $C_s^1$, $C_s^2$, $\ldots, C_s^{|C_s|}$, each with exactly one edge in $C_s$ (Line~\ref{line:kmcacc-partition}). We then construct $|C_s|$ number of $k$-MCA-CC problem instances, each with a new graph $G_i = (V_i, E_i)$, where $V_i = V$, $E_i = E \setminus C_s \cup C_s^i$. We recursively  solve $k$-MCA-CC on each graph $G_i$ to get $J_i$ (Line~\ref{line:kmcacc-recurse}).
Let $c(J)$ be the objective function in Equation~\eqref{eqn:kmcacc_obj}, the $J_i$ that minimizes the cost function,  $J^* = \argmin_{J_i}{c(J_i)}$, is the optimal solution to the original $k$-MCA-cc problem on $G$ (Line~\ref{line:kmcacc-argmin}). 

Note that our recursive call (Line~\ref{line:kmcacc-recurse}) partitions and reduces the solution space into disjoint subsets without losing optimal solutions, while the pruning step (Line~\ref{line:kmcacc-check-cost}) quickly prunes away unpromising solution spaces  leveraging efficient $k$-MCA algorithms. This two steps can conceptually be seen as a form of branch-and-bound used in the optimization literature~\cite{branch-and-bound}, but is specifically tailored to our graph problem (leveraging the mutually-exclusive nature of conflicting edges based on the FK-once property).

\begin{small}
\begin{algorithm}[t]
\SetKw{kwReturn}{return}
 \Input{Graph $G=(V, E)$}
 \Output{optimal solution to k-MCA-CC (k-snowflakes)}
 
$J \leftarrow$ Solve-k-MCA($G$) using Algorithm~\ref{alg:k-mca} \label{line:kmcacc-solve-kmca} 

  \uIf{$J$ is a feasible solution to $k$-MCA-CC(G)) \label{line:kmcacc-check}}{
    \kwReturn $J$ \label{line:kmcacc-return-early}
  }
  \uElseIf{  cost $c(J)$ is higher than best $k$-MCA-CC found so far \label{line:kmcacc-check-cost} }{
    \kwReturn null \label{line:kmcacc-return-null}
  }
  \uElse{
    $C_s \leftarrow$ edges in $J$ with the same source column index $s$ that violates FK-once constraint (Equation~\ref{eqn:kmcacc_fk_once} \label{line:kmcacc-else}) 

    $C_s^1$, $C_s^2$, \ldots $C_s^{|C_s|} \leftarrow$ disjoint subsets of $C_s$, each with exactly one edge from $C_s$ \label{line:kmcacc-partition}

    $E_i = E \setminus C_s \cup C_s^i, \forall i \in [|C_s|]$

    $G_i = (V, E_i)$ $\forall i \in [|C_s|]$

    $J_i = $ Call Solve-k-MCA-CC$(G_i)$ recursively, $\forall  i \in [|C_s|]$
\label{line:kmcacc-recurse}

    $J^* = \argmin_{J_i, i \in [|C_s|]}{c(J_i)}$ \label{line:kmcacc-argmin}
  }

\kwReturn $J^*$
\caption{Solve $k$-MCA-CC}
\label{alg:k-mca-cc}
\end{algorithm}
\end{small}

\begin{theorem}
\label{thm:optimality}
Algorithm~\ref{alg:k-mca-cc} solves $k$-MCA-CC optimally. 
\end{theorem}

\iftoggle{fullversion}
{
A proof of the theorem can be found in 
Appendix~\ref{apx:proof-optimality}.
}
{
A proof of the theorem can be found in 
our technical report~\cite{full}.
}
Intuitively, this algorithm ensures optimality, because at most one edge in $C_s$ may appear in the optimal solution to  $k$-MCA-CC (otherwise the FK-once constraint is violated), making edges in $C_s$ mutually exclusive. This  allows us to partition the conflicting edges in $C_s$, and recursively construct instances of the problem with smaller feasible regions, without losing the optimal solution in the process.

Given the hardness of $k$-MCA-CC, we clearly cannot hope Algorithm~\ref{alg:k-mca-cc} to solve $k$-MCA-CC in polynomial time. In practice, however, our branch-and-bound partitioning exploits the sparseness of edges (there are few edges pointing from the same columns), which turns out to be highly efficient. In our experiments (Section~\ref{sec:exp}), we observe the mean and median latency of Algorithm~\ref{alg:k-mca-cc}  on real-world BI schemas is 0.11 and 0.02 second, respectively, with the max being 11 seconds on a case with 88 data-tables (the largest case we encounter in the 100K+ real BI models harvested). We note that this is encouraging, and analogous to many classical combinatorial problems (e.g., Euclidean TSP and Knapsack) that are intractable in theory but reliably solvable in practice. 

\vspace{-0.5em}
\subsubsection{\textbf{Recall Mode (MaxEdge-CC)}} \hfill\\
\label{subsec:abi-r}
At this point, we did with the precision-mode of \abi, which is precision-oriented as we focus on finding the  salient $k$-snowflakes to reveal the ``backbone'' of the underlying schema.
However, not all desired joins are included in $k$-snowflakes/arboresence. For example, the green edges in Figure~\ref{fig:join-graph-constellation} correspond to the optimal solution to $k$-MCA-CC,  but we are still missing two joins in the ground truth, marked as dotted orange edges (these are joins that reference the same dimension-table, from the multiple fact tables). 

We, therefore, introduce the ``recall-mode'' of \abi, where we ``grow'' additional edges on top of the ``backbone'' identified in the precision-mode.

Specifically, given the original graph $G=(V, E)$, let $J^*$ be the optimal solution to $k$-MCA-CC. Let  $R = \{e_{ij} | e_{ij} \in (E \setminus J^*), P(e_{ij}) \geq \tau \}$ be the remaining edges that are promising (meeting a precision threshold $\tau$\footnote{By default, we threshold with $\tau$= 0.5 here, since our $P(e_{ij})$ are calibrated probability, and 0.5 is a natural cutoff for joinability.}) but not yet selected by $J^*$. For our recall-oriented solution of \abi, we want to select as many as edges $S \subseteq R$, subject to certain graph-structure constraints below. We write this as  EMS (Edge-Maximizing Schema) below:
\begin{align}
 \text{(EMS)} \qquad{}
\argmax_{S \subseteq R} &~~ |S| \label{eqn:ems_obj}  \\  
\mbox{s.t.} ~~ &  i \neq l, \forall e_{ij}, e_{lm} \in S  \cup J^*, j \neq m  \label{eqn:ems_fk_once}  \\
~~ &   S \cup J^* \text{is cycel-free}
\label{eqn:ems_no_cycles} 
\end{align} 


The new constraint in Equation~\eqref{eqn:ems_no_cycles} ensures that there should be no cycles in the resulting graph induced by $S \cup J^*$, as circular joins between columns are uncommon at a schema level.


The EMS problem is  NP-hard, and is 1/2-inapproximable using a reduction from Max-Sub-DAG~\cite{max-sub-dag}. However, because this step operates on top of the optimal solution $J^*$  to $k$-MCA-CC, there is typically limited flexibility in selecting from $R$. We find different solutions have very similar results (Section~\ref{sec:exp}), and we thus solve EMS greedily by picking the most confident edges (based on $P(e_{ij})$), without using more expensive solutions.  

This now completes the overall \abi. To recap, we first solve $k$-MCA-CC optimally using Algorithm~\ref{alg:k-mca-cc}  in the precision-mode, to get $J^*$ (the salient $k$-snowflakes). Then in the recall-mode, we solve EMS using $J^*$ and obtain $S^*$. The sub-graph induced by $J^* \cup S^*$ is the final solution of \abi.

\section{Experiments}
\label{sec:exp}

We perform large-scale evaluations to test \abi and related methods in the literature, in both quality and efficiency.



\iftoggle{fullversion}
{
\begin{table}[]
\footnotesize
    \centering
    \begin{tabular}{|c|c|c|c|c|}
    \hline
    &Average & 50-th p\% & 90-th p\%  & 95-th p\%  \\ \hline
    $\#$ of rows per table & 1053.4 & 50.1 & 1925.4 & 13981.2  \\ \hline
    $\#$ of columns per table & 8.1 & 4.2 & 11 & 17.1 \\ \hline
    $\#$ of tables (nodes) per case & 3.2 & 2 & 6.9 & 9.1 \\ \hline    $\#$ of relationships (edges) per case & 3.9 & 3 & 8.2 & 11.6\\ \hline 
    \end{tabular}
 \caption{Characteristics of all BI models harvested}. 
    \label{tab:stats_all}
\end{table}
}
{
}

\begin{table}[]
\footnotesize
    \centering
    \begin{tabular}{|c|c|c|c|c|}
    \hline
    &Average & 50-th p\% & 90-th p\%  & 95-th p\%  \\ \hline
    $\#$ of rows per table & 7730 & 80 & 10992 & 33125  \\ \hline
    $\#$ of columns per table & 9 & 5 & 20 & 27 \\ \hline
    $\#$ of tables (nodes) per case & 12.9 & 9 & 25 & 32 \\ \hline    $\#$ of relationships (edges) per case & 11.7 & 8.6 & 19 & 29\\ \hline 
    \end{tabular}
 \caption{Characteristics of 1000-case \textsc{Real} benchmark}. 
    \label{tab:stats_real}
\end{table}

\iftoggle{fullversion}
{
    \begin{table}[]
    \footnotesize
        \centering
        \begin{tabular}{|c|c|c|c|c|}
        \hline
        &TPC-C & TPC-E & TPC-H & TPC-DS  \\ \hline
        average $\#$ of rows per table & 288590 & 7850832 & 1082655 &  814889 \\ \hline
        average $\#$ of columns per table & 10.2  & 5.8 & 7.6 & 17.7 \\  \hline
        $\#$ of tables (nodes) &  9 & 32 & 8 & 24 \\ \hline
        $\#$ of relationships (edges) & 10 &45 &8 & 107 \\ \hline
        \end{tabular}
     \caption{Characteristics of 4 TPC benchmarks}. 
        \label{tab:stats_tpc}
    \end{table}
}
{
}

\begin{table*}[]
\vspace{-14mm}
\scriptsize
    \centering
    \scalebox{0.92}{
    \begin{tabular}{|c|l||c|c|c|c||c|c|c||c|c|c||c|c|c||c|c|c||}
    \hline
    & & \multicolumn{4}{c||}{Real (OLAP)} & \multicolumn{6}{c||}{Synthetic (OLAP)} & \multicolumn{6}{c||}{Synthetic (OLTP)}  \\ \hline
     &Benchmark &\multicolumn{4}{c||}{1000-case \textsc{Real} benchmark} & \multicolumn{3}{c||}{TPC-H} & \multicolumn{3}{c||}{TPC-DS} & \multicolumn{3}{c||}{TPC-C} & \multicolumn{3}{c||}{TPC-E}\\ \hline
     Method Category & Method & $P_{edge}$ & $R_{edge}$ &  $F_{edge}$ & $P_{case}$ & $P_{edge}$ & $R_{edge}$ &  $F_{edge}$  & $P_{edge}$ & $R_{edge}$ &  $F_{edge}$ & $P_{edge}$ & $R_{edge}$ &  $F_{edge}$ & $P_{edge}$ & $R_{edge}$ &  $F_{edge}$\\ \hline
    \multirow{3}{4em}{\abi}&\abi-P & \textbf{0.98} & 0.664 &  0.752 & \textbf{0.92} & \textbf{1} & 0.88& 0.93  & \textbf{0.99} & 0.28& 0.43 & \textbf{1} & 0.6& 0.75 & \textbf{1} & 0.4& 0.58\\ 
    &\abi & \textbf{0.973} &  \textbf{0.879} &  \textbf{0.907} & 0.853 & \textbf{1} & \textbf{1}& \textbf{1} & 0.96 & \textbf{0.91}& \textbf{0.93} & \textbf{1} & \textbf{0.8}& \textbf{0.89}& 0.96 & \textbf{0.93}& \textbf{0.95}\\
    &\abi-S & 0.951 &  0.848 &  0.861 & 0.779 & \textbf{1} & \textbf{1} & \textbf{1} & 0.92 & 0.89 & 0.91 & \textbf{1} & 0.7 & 0.82& 0.93 & \textbf{0.94} & \textbf{0.94}\\ \hline
    Commercial & \pbi & 0.916 & 0.584 & 0.66 & 0.754 &0 & 0 & 0 &0 & 0 & 0& 0 & 0 & 0 & 0 & 0 & 0 \\ 
    \hline
    \multirow{4}{4em}{Baselines}& MC-FK & 0.604 & 0.616 & 0.503 & 0.289 & \textbf{1} & \textbf{1}&\textbf{1}  & 0.73 & 0.65 & 0.68 & 0.46 & \textbf{0.8} & 0.63 & 0.57 & 0.79 & 0.48 \\ 
    &Fast-FK & 0.647 & 0.585 & 0.594 & 0.259 & 0.71 & 0.88 & 0.79& 0.62 & 0.35 & 0.44 & 0.62 & 0.57 & 0.6& 0.73 & 0.84 & 0.78\\ 
    &HoPF & 0.684 & 0.714 & 0.67 & 0.301 &0.86 &0.75 &0.8&0.87 &0.51 &0.65 &0.75 &0.7 &0.72 &0.71 &0.91 &0.81\\ 
    &\textcolor{black}{ML-FK} & \textcolor{black}{0.846} & \textcolor{black}{0.77} & \textcolor{black}{0.773} & \textcolor{black}{0.557} & \textcolor{black}{0.6}  & \textcolor{black}{0.75} & \textcolor{black}{0.667} & \textcolor{black}{0.369} & \textcolor{black}{0.589} & \textcolor{black}{0.454} & \textcolor{black}{0.273}  & \textcolor{black}{0.3}  & \textcolor{black}{0.286} & \textcolor{black}{0.694} & \textcolor{black}{0.756} & \textcolor{black}{0.723} \\  \hline
    {Language model} &GPT-3.5 & 0.73 & 0.64 & 0.67 & 0.43 & 0.75 & 0.75 & 0.75 & 0 & 0& 0 & 0.16 & 0.2 & 0.15 & 0.78 & 0.56 & 0.65 \\ 
    \hline
    \end{tabular}}
    \caption{Quality comparison on the 1000-case \textsc{Real} benchmark and 4 TPC benchmarks. }
    \vspace{-1em}
    \label{tab:quality}
\end{table*}

\iftoggle{fullversion}
{
    \begin{table*}[]
    \scriptsize
        \centering
        \scalebox{0.85}{
        \begin{tabular}{|c|l||c|c|c||c|c|c||c|c|c||c|c|c||c|c|c||c|c|c||c|c|c||c|c|c||}
        \hline
        & & \multicolumn{12}{c||}{Denormalized (OLAP-like)} & \multicolumn{12}{c||}{Normalized (OLTP-like)}  \\ \hline
         &Benchmark &\multicolumn{3}{c||}{Foodmart} & \multicolumn{3}{c||}{Northwind} & \multicolumn{3}{c||}{AdventureWork} & \multicolumn{3}{c||}{WorldWideImporters} & \multicolumn{3}{c||}{Foodmart} & \multicolumn{3}{c||}{Northwind} & \multicolumn{3}{c||}{AdventureWork} & \multicolumn{3}{c||}{WorldWideImporters} \\  \hline
         Method Category & Method & $P_e$ & $R_e$ &  $F_e$ & $P_e$ & $R_e$ &  $F_e$  & $P_e$ & $R_e$ &  $F_e$ & $P_e$ & $R_e$ &  $F_e$ & $P_e$ & $R_e$ &  $F_e$ & $P_e$ & $R_e$ &  $F_e$  & $P_e$ & $R_e$ &  $F_e$ & $P_e$ & $R_e$ &  $F_e$\\ \hline
        \multirow{3}{4em}{\abi}&\abi-P & \textbf{1} & 0.5 & 0.67 & \textbf{1} & \textbf{1} & \textbf{1} & \textbf{1} & 0.34 & 0.51 & \textbf{1} & 0.31 & 0.47 & \textbf{1} & 0.63 & 0.77 & \textbf{1} & \textbf{1} & \textbf{1} & \textbf{1} & 0.44 & 0.62 & \textbf{1} & 0.25 & 0.4 \\ 
        & \abi & 0.75 & \textbf{1} & \textbf{0.86} & \textbf{1} & \textbf{1} & \textbf{1} & 0.98 & \textbf{0.93} & \textbf{0.97} & \textbf{1} & 0.83 & \textbf{0.91} & 0.8 & \textbf{1} & \textbf{0.89} & \textbf{1} & \textbf{1} & \textbf{1} & 0.97 & 0.82 & \textbf{0.89} & 0.97 & \textbf{0.86} & \textbf{0.91} \\ 
        & \abi-S & 0.68 & \textbf{1} & 0.81 & 0.9 & \textbf{1} & 0.95 & 0.94 & 0.91 & 0.92 & \textbf{1} & 0.79 & 0.88 & 0.8 & \textbf{1} & 0.88 & \textbf{1} & \textbf{1} & \textbf{1} & 0.93 & 0.8 & 0.86 & 0.92 & 0.84 & 0.88 \\
        \hline
        Commercial & \pbi & 0.75 & 0.6 & 0.67 & 0.85 & 0.79 & 0.82 & 0.97 & 0.66 & 0.78 & \textbf{1} & 0.59 & 0.74 & 0.75 & 0.6 & 0.67 & 0.9 & \textbf{1} & 0.95 & 0.9 & 0.34 & 0.5 & \textbf{1} & 0.09 & 0.17  \\
        \hline
        \multirow{5}{4em}{Baselines}& MC-FK & 0.42 & 0.9 & 0.57 & 0.28 & \textbf{1} & 0.44 &  0.33 & 0.71 & 0.45 & 0.21 & \textbf{0.86} & 0.34 & 0.54 & 0.88 & 0.67 & 0.43 & 0.8 & 0.56 & 0.2 & \textbf{0.87} & 0.33 & 0.2 & 0.52 & 0.3\\ 
        & Fast-FK & 0.38 & 0.6 & 0.46 & 0.53 & 0.64 & 0.65 & 0.33 & 0.34 & 0.34 &  0.27 & 0.47 & 0.35 & 0.55 & 0.75 & 0.63 & 0.55 & 0.6 & 0.57 & 0.43 & 0.69 & 0.53 & 0.45 & 0.14 & 0.2 \\
        & HoPF & 0.44 & 0.4 & 0.42 & 0.53 & 0.5 & 0.52 &0.78 & 0.73 & 0.75 & 0.89 & 0.72 & 0.8  & 0.43 & 0.38 & 0.4 & 0.8 & 0.8 & 0.8 & 0.55 & 0.71 & 0.62 & 0.75 & 0.51 & 0.61\\ 
        & ML-FK & 0.43 & 0.8 & 0.56 & 0.35 & 0.86 & 0.5 & 0.84 & 0.8 & 0.82 & 0.89 & 0.75 & 0.81 & 0.57 & 0.88 & 0.69 & 0.5 & 0.9 & 0.64 & 0.95 & 0.83 & \textbf{0.89} & 0.9 & 0.5  & 0.64 \\ \hline
        {Language model} &GPT-3.5 & \textbf{1} & 0.6 & 0.75 & \textbf{1} & 0.71 & 0.83 & \textbf{1} & 0.32 & 0.48 & 0.77 & 0.69 & 0.73 & \textbf{1} &  0.75 & 0.86 & \textbf{1} & \textbf{1} & \textbf{1} & 0.87 & 0.54 & 0.67 & 0.88 & 0.61 & 0.72 \\ 
        \hline
        \end{tabular}}
        \caption{\small Quality comparison on FoodMart, NorthWind, AdventureWork and WorldWideImporters . ($P_e,R_e,F_e$) = ($P_{edge},R_{edge},F_{edge}$). }
        \vspace{-1em}
        \label{tab:new-synthetic-quality}
    \end{table*}
}
{

}

    

\subsection{Evaluation Setup}

\begin{table*}[]
\scriptsize
    \centering
    \scalebox{0.92}{
    \begin{tabular}{|c|c|c|c|c|c|c|c|c|c|c|c|}
    \hline
    & \# of tables  & 4 & 5 & 6 & 7 & 8 & 9 & 10 & [11,15] & [16,20] & 21+ \\ \hline
\textcolor{black}{Case-type statistics} & \textcolor{black}{(ST,SN,C,O)} & \textcolor{black}{(50,18,31,1)} & \textcolor{black}{(49,12,37,2)} & \textcolor{black}{(36,14,48,2)} & \textcolor{black}{(19,23,49,9)} &  \textcolor{black}{(10,22,57,11)} & \textcolor{black}{(7,25,50,18)} & \textcolor{black}{(2,39,40,19)} & \textcolor{black}{(7,14,60,19)} & \textcolor{black}{(1,7,72,20)} & \textcolor{black}{(9,5,62,24)} \\  
\hhline{|=|=|=|=|=|=|=|=|=|=|=|=|}
\multirow{3}{4em}{\abi}& \abi &\textbf{0.97} (0.99,0.95)&\textbf{0.97} (0.98,0.96)&\textbf{0.96} (0.99,0.94)&\textbf{0.95} (0.98,0.91)&\textbf{0.95} (0.99,0.92)&\textbf{0.96} (1.00,0.93)&\textbf{0.94} (0.98,0.90)&\textbf{0.90} (0.97,0.85)&\textbf{0.84} (0.95,0.75)&\textbf{0.79} (0.94,0.69)\\ 
&\abi-P &0.91 (1.00,0.84)&0.91 (0.99,0.85)&0.88 (1.00,0.79)&0.81 (0.98,0.69)&0.83 (0.98,0.71)&0.83 (1.00,0.70)&0.81 (0.99,0.69)&0.71 (0.98,0.56)&0.60 (0.96,0.44)&0.55 (0.95,0.39)\\ 
&\abi-S & 0.95 (0.98,0.93)&0.95 (0.96,0.94)&0.94 (0.97,0.92)&0.94 (0.97,0.90)&0.95 (0.98,0.91)&0.93 (0.97,0.89)&0.92 (0.96,0.88)&0.85 (0.92,0.79)&0.78 (0.90,0.70)&0.74 (0.89,0.63)\\ \hline
Commercial & \pbi &0.76 (0.94,0.66)&0.67 (0.91,0.55)&0.76 (0.94,0.66)&0.76 (0.93,0.66)&0.75 (0.91,0.65)&0.78 (0.91,0.70)&0.77 (0.92,0.67)&0.74 (0.90,0.65)&0.65 (0.80,0.56)&0.66 (0.88,0.54)\\ 
\hline
\multirow{4}{4em}{Baselines}& MC-FK & 0.69 (0.88,0.57)&0.65 (0.93,0.49)&0.63 (0.70,0.58)&0.65 (0.67,0.63)&0.62 (0.70,0.56)&0.56 (0.46,0.72)&0.54 (0.48,0.63)&0.54 (0.49,0.61)&0.52 (0.42,0.69)&0.42 (0.30,0.68)\\ 
&Fast-FK &0.76 (0.79,0.72)&0.76 (0.79,0.74)&0.68 (0.69,0.66)&0.53 (0.53,0.52)&0.65 (0.67,0.63)&0.47 (0.46,0.47)&0.45 (0.48,0.42)&0.49 (0.53,0.46)&0.47 (0.49,0.46)&0.42 (0.40,0.44)\\ 
&HoPF &0.83 (0.86,0.81)&0.77 (0.77,0.76)&0.72 (0.73,0.71)&0.63 (0.58,0.70)&0.68 (0.67,0.70)&0.55 (0.49,0.64)&0.62 (0.55,0.70)&0.61 (0.58,0.64)&0.57 (0.55,0.60)&0.49 (0.44,0.56)\\ 
&\textcolor{black}{ML-FK} & \textcolor{black}{0.87(0.91,0.84)} & \textcolor{black}{0.86 (0.91,0.85)} & \textcolor{black}{0.8 (0.94,0.79)} & \textcolor{black}{0.83 (0.88,0.83)} & \textcolor{black}{0.84 (0.89,0.84)} & \textcolor{black}{0.8 (0.84,0.81)} & \textcolor{black}{0.8 (0.88,0.79)} & \textcolor{black}{0.72 (0.84,0.71)} & \textcolor{black}{0.64 (0.69,0.65)} & \textcolor{black}{0.56 (0.65,0.6)} \\ \hline
{Language model} &GPT-3.5 & 0.79 (0.86,0.76) & 0.75 (0.83,0.7) & 0.8 (0.83,0.78) & 0.74 (0.78,0.74) & 0.72 (0.77,0.69) & 0.69 (0.76,0.66) & 0.69 (0.76,0.66) & 0.56 (0.62,0.54) & 0.49 (0.56,0.46) & 0.39 (0.46,0.36) \\ 
\hline
    \end{tabular}
    }
    \caption{\small Edge-level quality reported as ``F-1 (precision, recall)'',  by number of input tables in \textsc{Real} benchmark. In the first row, we report ``case type statistics'', denoted as ($ST,SN,C,O$), which stand for the number of cases in (Star, Snowflake, Constellation,  Others), respectively.}
    \vspace{-1.5em}
    \label{tab:pr_result}
\end{table*}

\textbf{Benchmarks.} We use two benchmarks to evaluate \abi.

- \textbf{\textsc{Real}}. 
We sample 1000 real BI models crawled in the wild (Section~\ref{subsec:data}) as our first benchmark, henceforth referred to as {\textsc{Real}}.\footnote{To be released on GitHub after an internal review~\cite{code}}. In order to account for different levels of difficulty in predicting BI models, we perform stratified sampling as follows -- we bucketize models into 10 groups based on the number of input tables as \{4, 5, 6, 7, 8, 9, 10, [11-15], [16-20], 21+\}, and randomly select 100 cases in each group, to create this 1000-case benchmark that covers the entire spectrum in terms of levels of difficulty, from easy (with only a few tables) to hard (with over 20 tables). Table~\ref{tab:stats_real} summarizes the characteristics of the 1000-case {\textsc{Real}} benchmark. Note that these test cases are held out and never used when training of our local classifier, which is consistent with the standard practice of machine learning to avoid data leakage.

\iftoggle{fullversion}
{
In comparison, Table~\ref{tab:stats_all} shows the characteristics of all BI models harvested in the entire collection, which are substantially simpler (e.g., with a small number of tables). This is not entirely surprising, when these BI models are programmed by non-technical business users using GUI tools, which is why we perform stratified sampling described above to create a more balanced and challenging test set.
}
{
}

- \textbf{\textsc{Synthetic}}. In addition to evaluation on real BI models, we perform tests on 4 TPC benchmarks, henceforth referred to as \textsc{Synthetic}. 
Specifically, {TPC-H} and {TPC-DS} are two popular benchmarks for BI/OLAP workloads. We evaluate predicted joins with the ground-truth relationships in TPC specifications. We further perform tests on TPC-C and {TPC-E}, two popular OLTP benchmarks. While \abi is not designed for general OLTP databases, we perform the tests nevertheless, in order to understand \abi's ability to detect foreign-keys in general (beyond the snowflake-like schemas that \abi was initially designed for). 

We further perform tests on 4 commonly-used synthetic databases: FoodMart~\cite{foodmart}, Northwind~\cite{northwind}, AdventureWorks~\cite{adventurework} and WorldWide-Importers~\cite{worldwideimporters}. We use both of their OLAP  and OLTP versions to stress test our algorithms, for a total of 8 additional test databases. 
\iftoggle{fullversion}
{
}
{
We report these additional results in our technical report~\cite{full} in the interest of space. 
}

\begin{table}[]
\small
    \centering
    \scalebox{0.67}{
    \begin{tabular}{|c|c|c|c|c|c|c|c|c|c|c|c|}
    \hline
    &\# of tables  & 4 & 5 & 6 & 7 & 8 & 9 & 10 & [11,15] & [16,20] & 21+ \\ \hline
\multirow{3}{4em}{\abi}& \abi &\textbf{1.00}&0.96&0.95&0.89&0.95&0.97&0.85&0.78&0.64&0.55\\    
&\abi-P &\textbf{1.00}&\textbf{0.98}&\textbf{0.99}&\textbf{0.94}&\textbf{0.96}&\textbf{0.99}&\textbf{0.95}&\textbf{0.89}&\textbf{0.83}&\textbf{0.67}\\ 
&\abi-S & 0.99&0.95&0.93&0.93&0.95&0.80&0.76&0.69&0.49&0.31\\ \hline
Commercial & \pbi &0.91&0.87&0.81&0.85&0.77&0.78&0.74&0.75&0.52&0.54\\  
\hline
\multirow{5}{4em}{Baselines}&MC-FK &0.76&0.68&0.41&0.34&0.19&0.14&0.1&0.13&0.09&0.05\\ 
&Fast-FK &0.68&0.65&0.39&0.14&0.32&0.08&0.09&0.12&0.09&0.03\\ 
&HoPF &0.78&0.57&0.42&0.21&0.32&0.06&0.18&0.19&0.18&0.11\\
&\textcolor{black}{ML-FK} & \textcolor{black}{0.87} & \textcolor{black}{0.86} & \textcolor{black}{0.68} & \textcolor{black}{0.65} & \textcolor{black}{0.7} & \textcolor{black}{0.55} & \textcolor{black}{0.53} & \textcolor{black}{0.42} & \textcolor{black}{0.18} &\textcolor{black}{0.12}
\\ \hline
{Language model} &GPT-3.5 & 0.67 & 0.61 & 0.61 & 0.6 & 0.44 & 0.42 & 0.42 & 0.21 & 0.1 & 0.05 \\\hline

    \end{tabular}
    }
    \caption{Case-level precision, by number of tables.}
    \label{tab:pc_result}
\end{table}

\textbf{Metrics.} We compare the predicted joins of different methods, against the ground truth (which in the case of \textsc{Real}, are human-specified relationships that we programmatically extract from BI models we crawled). 
We evaluate prediction quality of algorithms both at the \textit{edge-level} and \textit{case-level}, defined as below:

\underline{Edge-level quality}. For each BI test case $C$, we evaluate the fraction of predicted join relationships (edges on the graph) that is identical to ground-truth relationships.  We use standard precision/recall/F metrics for the edge-level evaluation, defined as precision $P_{edge}(C) = \frac{\text{num-of-correctly-predicted-edges}}{\text{num-of-total-predicted-edges}}$, recall $R_{edge}(C) = \frac{\text{num-of-correctly-predicted-edges}}{\text{num-of-total-ground-truth-edges}}$. The F-1 score, denoted by $F_{edge}(C)$, is then the harmonic mean of precision and recall, or $\frac{2 P_{edge}(C) R_{edge}(C)}{P_{edge}(C) + R_{edge}(C)}$. 

We report precision/recall/F-1 for the entire \textsc{Real} benchmark, as the average across all 1000 test cases. 

\underline{Case-level quality}.
Since it is difficult for non-technical users to identify incorrectly-predicted relationships,  high precision is crucial for \abi (Section~\ref{subsec:abi-p}). We report \textit{case-level} precision for each test case $C$, denoted by $P_{case}(C)$, defined as:
  \begin{equation}
    P_{case}(C)=
    \begin{cases}
      1, & \text{if}\ P_{edge}(C) = 1  \\
      0, & \text{otherwise}
    \end{cases}
  \end{equation}
In other words, $P_{case}(C)$ is 1 only if no incorrect edge is predicted; and 0 even if a single false-positive edge is predicted (for it is unlikely that non-technical users can identify and correct the incorrect predictions, making us to ``fail'' in such a situation). The case-level precision on the entire benchmark is also the average across all 1000 cases.\footnote{We note that it is possible that a predicted join graph may differ syntactically from the ground truth when the two are semantically equivalent. For example, the ground truth may have a fact-table $F$ join a dimension-table $A$ (N:1), which in turn 1:1 join with another dimension-table $B$. If it is predicted that $F$ joins with $B$ (N:1), and then $B$ joins with $A$ (1:1), while the ground truth is F-(N:1)-A-(1:1)-B, then the are in fact identical  (except that the join order is different). We account for semantic equivalence in our evaluation and mark both as correct, which is fair for all methods tested.}

\underline{Latency}. We report the latency of all methods using wall-clock time. All methods are implemented using Python 3.8 and tested on a Windows 11 machine with 64-core Intel CPU and 128 GB memory. 

\begin{figure*}
\vspace{-18mm}
\centering
\subfigure[50/90/95-th Percentiles]{
\label{fig:percentile}
\includegraphics[width=0.37\textwidth]{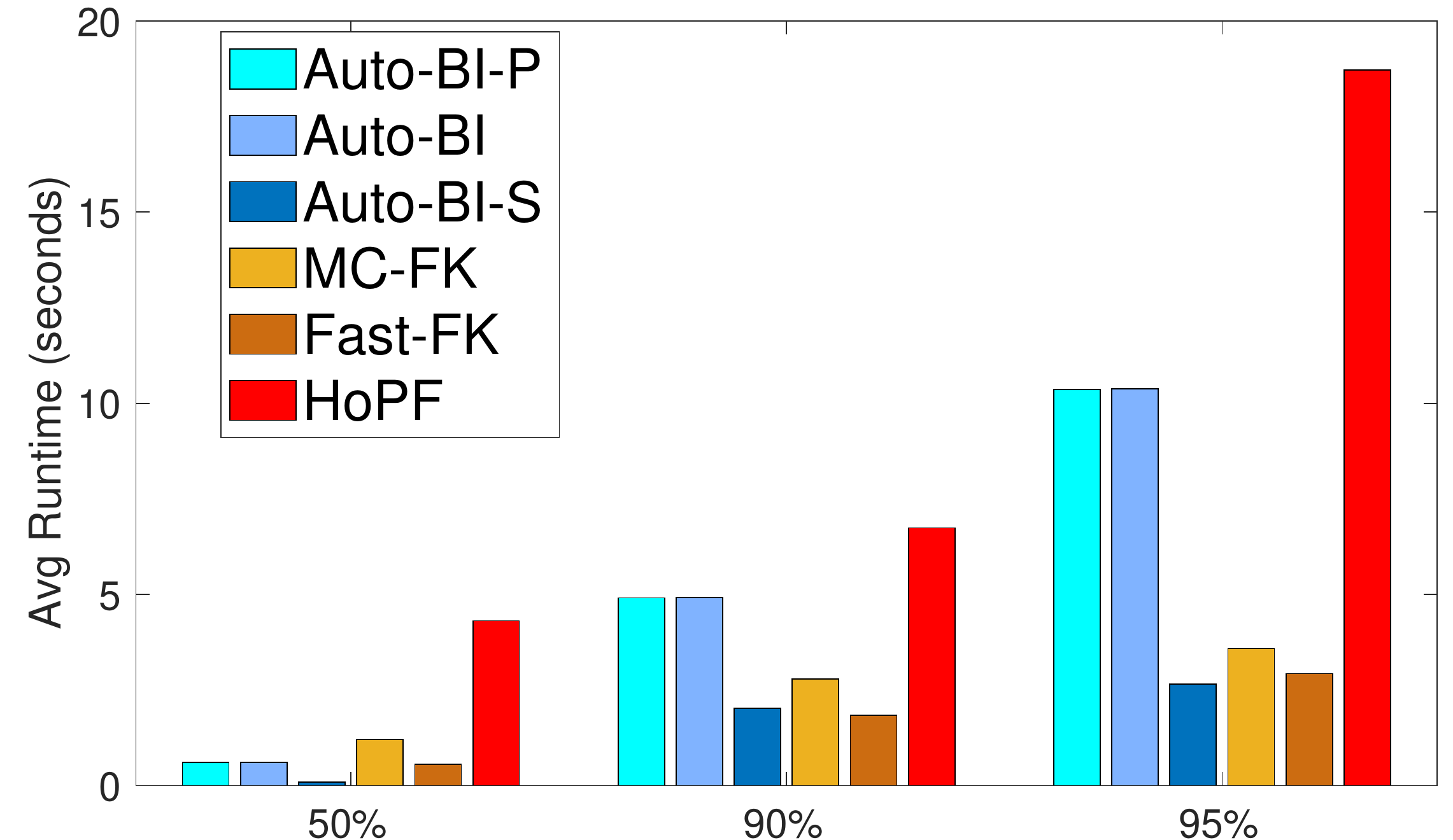}}
\hspace{25mm}
\subfigure[Breakdown by components]{
\label{fig:breakdown}
\includegraphics[width=0.37\textwidth]{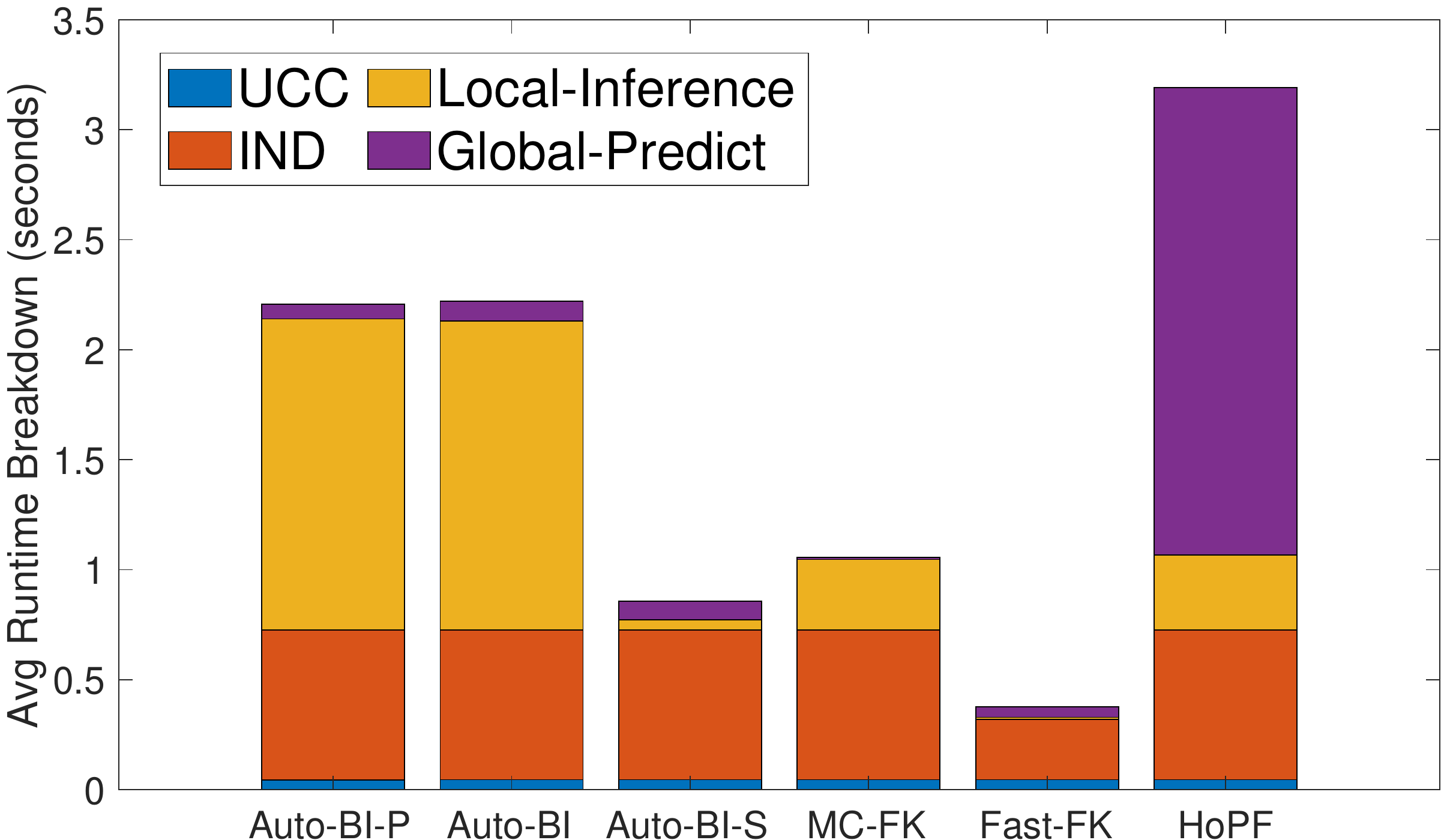}}
\caption{Comparison of end-to-end latency.}
\vspace{-1em}
\label{fig:end_to_end_time}
\end{figure*}

\subsection{Methods compared}
\label{subsec:methods}
\vspace{-1mm}
We compare \abi with the following  methods in the literature. 

\textbf{MC-FK~\cite{zhang2010multi}}. MC-FK pioneered a method to accurately detect multi-column FKs, using an EMD-based randomness metric that measures the distribution similarity of two columns. 

\textbf{Fast-FK~\cite{chen2014fast}}. Fast-FK makes FK predictions based on a scoring function that combines column-value and column-name similarity. This method employs fast pruning and focuses on ensuring  interactive predictions. 

\textbf{HoPF~\cite{hpi}}. HoPF is a recent method that detects FKs and PKs together, in order to ensure better prediction accuracy. It employs hand-tuned scoring functions as well as structural constraints (e.g., no cycles) to improve accuracy.  

\textbf{ML-FK~\cite{fk-ml}}. ML-FK is an ML-based approach to predict FK, which is similar in spirit to our local-classifiers, and we feed ML-FK with the same training data used by our local-classifiers. 

\textbf{\pbi}. \pbi is a commercial system from a leading BI vendor that has a feature to detect joins. We anonymize the name of the system,  in keeping with benchmarking traditions in the database literature~\cite{dewitt-clause, pavlo2009comparison, schmidt2002xmark, 
funke2011benchmarking}.

\textbf{GPT-3.5}. GPT is a family of  language models~\cite{brown2020language} capable of performing many downstream tasks such as NL-to-SQL~\cite{rajkumar2022evaluating, wang2022proton} and schema-matching~\cite{schemamatchingLLM}. We use GPT as a baseline with few-shot learning~\cite{brown2020language}, where we provide few-shot demonstrations of the join task in the prompt, and then ask the model to predict for new test cases. We use the latest publicly available version of GPT (GPT-3.5-turbo), and in the prompt we provide both the table-schema (column names) as well as sample data rows. Note that because synthetic benchmarks (e.g., TPC) are heavily represented on the web, it is likely that GPT has seen such cases in its pre-training (e.g.,  TPC queries on the web), causing data leakage and inflated results. We report these numbers nevertheless, for reference purposes.

\textbf{\abi}. This is our proposed method using global $k$-MCA optimization. We evaluate three variants of our method: (1): \textbf{\abi-Precision (\abi-P)} is the precision-mode of \abi (Section~\ref{subsec:abi-p}) that focuses on high precision (by finding snowflake-like ``backbones'' from the schema). (2): \textbf{\abi} is our full approach with both precision and recall modes (Section~\ref{subsec:abi-r}).  (3): Furthermore, we test a light-weight version of \abi called \textbf{\abi-Schema-Only (\abi-S)}, which uses only schema-level metadata (table/column names) for local-classifiers (without using column-values). This method is thus efficient, with sub-second latency in most cases, while still being surprisingly accurate.

\begin{figure}[tb]
	\centering
	\includegraphics[width=0.6\linewidth]{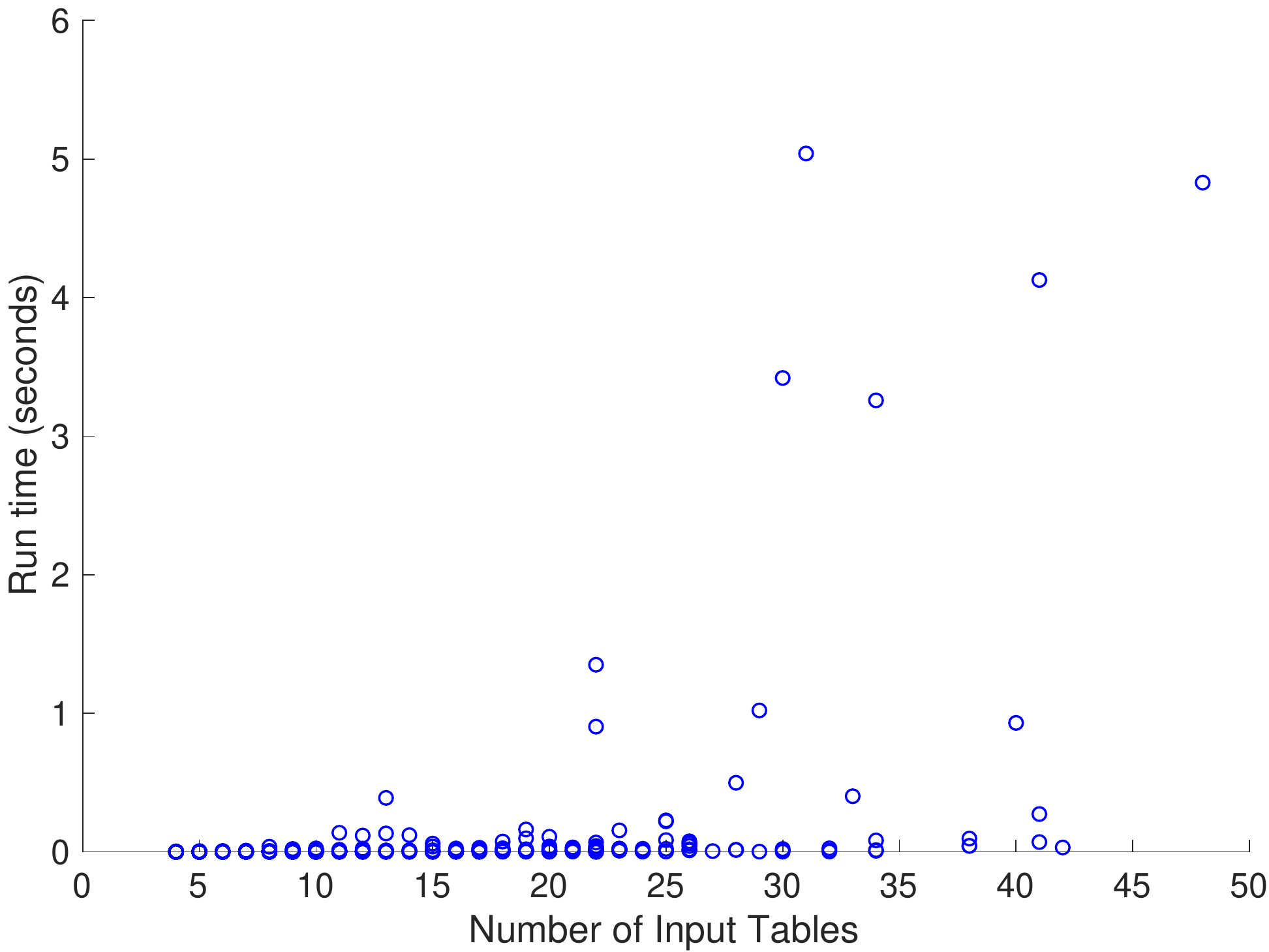}
	\caption{\small  Latency distribution of k-MCA-CC on 1000-case \textsc{Real}. The 50/90/95-th p-tiles are 0.02/0.06/0.17 seconds, respectively. 
  }
	\label{fig:k-mca-time-1000}
\end{figure}

\begin{figure*}
\vspace{-18mm}
	\centering
\begin{minipage}[t]{0.45\linewidth}
   \centering
{\includegraphics[width=0.9\linewidth]{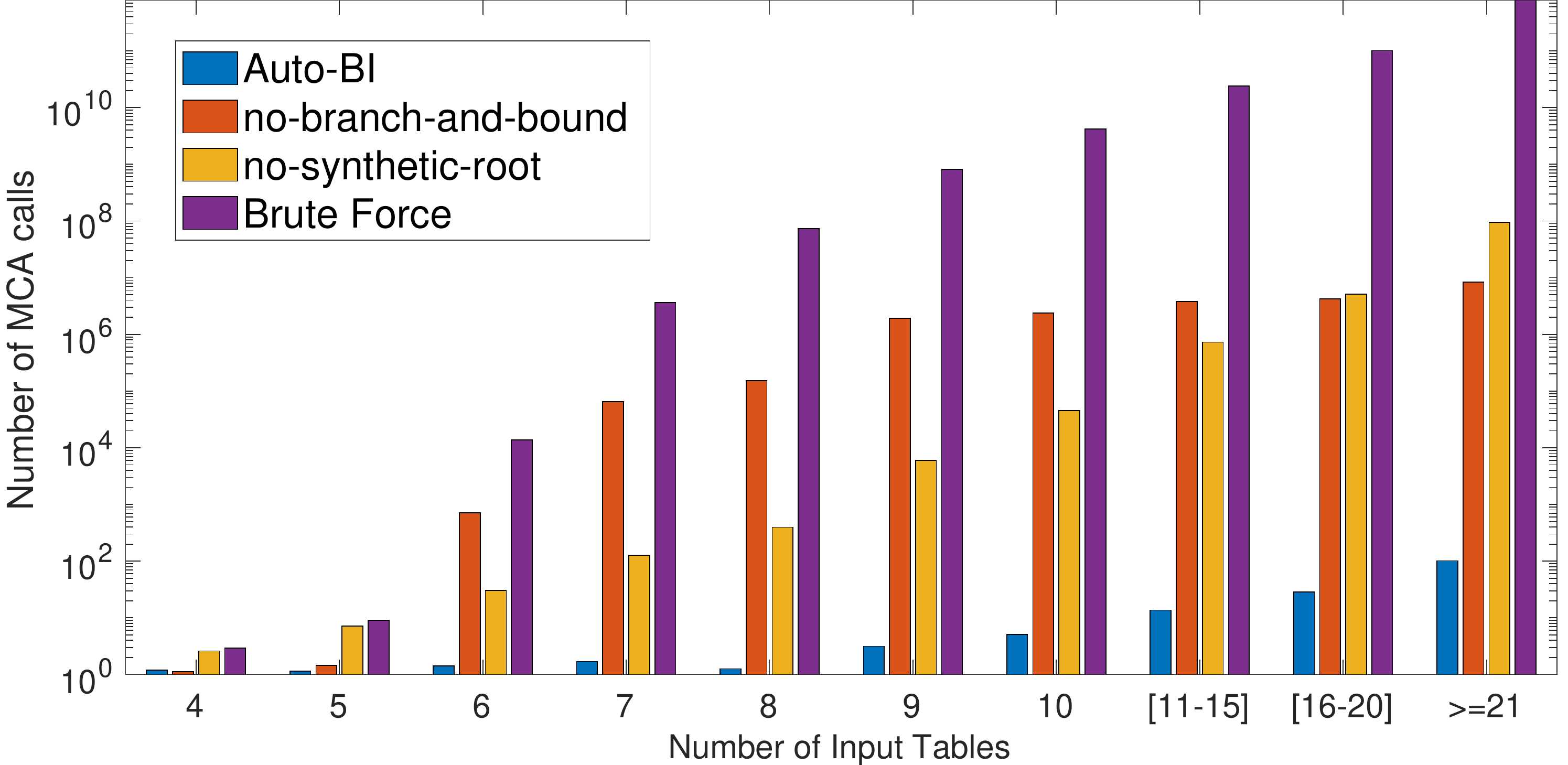}}
		\caption{\small Effect of our efficiency optimization techniques, as measured by the number of 1-MCA invocations.} 
	\label{fig:efficiency_ablation}
	\end{minipage}
 \hfill
  \begin{minipage}[t]{0.45\linewidth}
  \centering
{\includegraphics[width=0.9\linewidth]{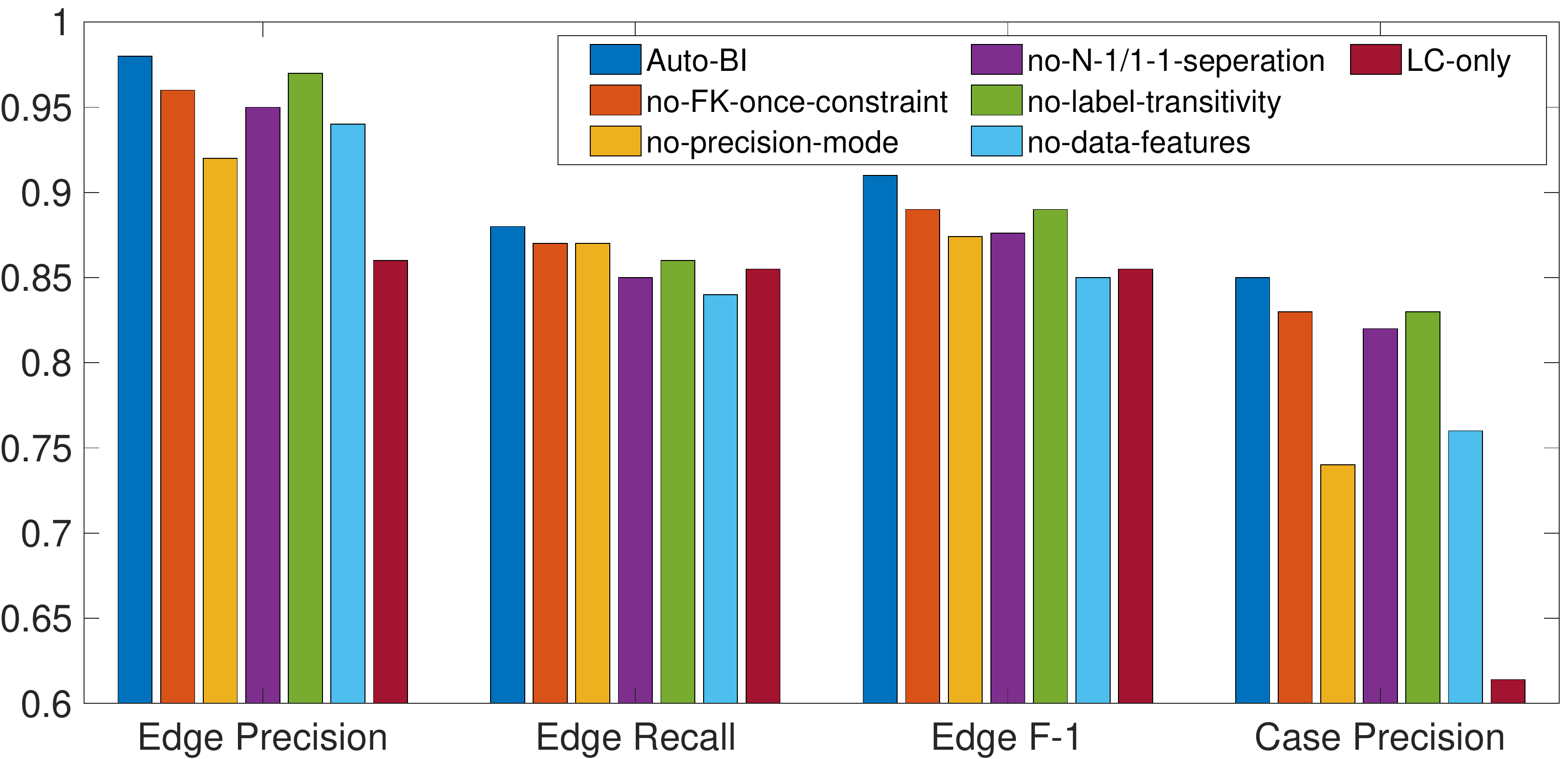}}
		\caption{\small Ablation study on the effect of \abi components on result quality (average results on the 1000-case \textsc{Real}). }
	\label{fig:ablation}
	\end{minipage}	
\vspace{-3mm}
\end{figure*}

\vspace{-1em}
\subsection{Quality comparisons}
\label{subsec:exp_overall}

\textbf{Overall comparison.} 
Table~\ref{tab:quality} compares the overall quality of all methods, on both the 1000-case \textsc{Real} benchmark and 4 \textsc{Synthetic} TPC benchmarks. As we can see, \abi-based methods consistently outperform alternatives, with \abi-P  always being the best in precision (very close to 1), while the full \abi being the best overall in F-scores. 

On the 1000-case \textsc{Real} benchmark, we note that the difference between \abi and the best baselines is significant -- the difference is  13 percentage points for edge-level precision (0.98 vs. 0.846), and over 10 percentage points for edge-level recall (0.879 vs. 0.77). The difference is even more pronounced at the case-level, which is over 35 percentage points (0.92 vs. 0.557), showing the strong benefit on quality when using a principled global optimization in our $k$-MCA. 

Even the light-weight \abi-S (with the same $k$-MCA but using schema-only features) is surprisingly accurate (losing only 2-3 percentage points in precision/recall). As we will see, it is much more efficient, however, making it a strong contender for practical use when latency becomes important.

For the \textsc{Synthetic} TPC benchmarks, we observe  similar trends  consistent with the  \textsc{Real} benchmark.  It is interesting to note that, not only is \abi the best on OLAP benchmarks (TPC-H/TPC-DS), it turns out to be the best on OLTP benchmarks too (TPC-C/TPC-E). This is somewhat surprising, because OLTP databases do not usually follow snowflake-like schema, and are not the use cases that \abi is originally designed for. We inspected TPC-C and TPC-E carefully to understand why \abi is effective. It turns out that while these OLTP databases do not have snowflake designs, they nevertheless have clusters of tables (e.g., a cluster of tables on ``\code{Customers}'', and another cluster on ``\code{Market}'', etc., for TPC-E), where each cluster loosely follows a hub-and-spoke pattern with related tables joining through a few ``central'' tables (e.g., a cluster of tables  relating to customers, such as  ``\code{Customers-Account}'' and ``\code{Customers-Tax-Rate}'', all join through a central ``\code{Customers}'' table in  TPC-E, like very nicely depicted in Figure 2 of prior work on schema summarization~\cite{yang2009summarizing}).  
\iftoggle{fullversion}
{
Similar results are also observed in additional synthetic benchmarks as shown in Table~\ref{tab:new-synthetic-quality}.
}
{
Similar results are also observed in additional synthetic benchmarks~\cite{full}.
}
Exploring the applicability of \abi-like global optimization in FK-detection for general OLTP databases is an interesting area for future work.

Among all baselines, the commercial \pbi produces high precision at the cost of a low recall. Among the four FK methods from the literature, ML-FK produces the best quality (since we feed it with the same training data harvested from real BI models), while the remaining three have comparable precision/recall. 


\textbf{Detailed breakdown.}
Table~\ref{tab:pr_result} and Table~\ref{tab:pc_result}  show a more detailed comparison of edge-level and case-level quality, respectively, bucketized by the number of input tables. 
The overall trend is consistent with Table~\ref{tab:quality} -- \abi has the best F-1, while \abi-P has the best precision across the board. We note that \abi works reasonably well on the most challenging slice of the test cases (with 21+ tables), maintaining high precision (0.94) and reasonable recall.






\iftoggle{fullversion}
{
We report additional experiments and analysis (comparing with baselines enhanced using components from \abi to better understand its component-wise benefits) in Appendix~\ref{apx:add-exp}.
}
{
We report additional experiments and analysis (comparing with baselines enhanced with components from \abi, to understand the contribution of \abi components) in~\cite{full}.
}


\vspace{-1em}
\subsection{Efficiency comparison}
\label{subsec:exp_efficiency}
\textbf{Overall comparison.}
We show the 50/90/95-th percentile latency of all methods in Figure~\ref{fig:end_to_end_time}(a), on the 1000-case \textsc{Real} benchmark.  Overall, \abi-S and Fast-FK are the fastest, taking 2-3 seconds even in the largest cases. \abi is 2-3x slower but the latency is still acceptable. HoPF takes the most time in comparison.

 

We also report the latency breakdown  of all methods in Figure~\ref{fig:end_to_end_time}(b). 
Overall, there are 4 main components for all methods:  (1) \textsf{UCC}: generate unique column combinations; (2) \textsf{IND}: generate inclusion dependencies; (3) \textsf{Local-Inference}: generate features and perform local-classifier inference for joins in \abi (while baselines use predefined scoring functions here); (4) \textsf{Global-Predict}: \abi uses $k$-MCA, whereas baselines use various other algorithms. 
For \abi{}, the most expensive part is \textsf{Local-Inference} (specifically, feature generation for inference). \abi-S  is effective in reducing the latency of this part (by using metadata-only features), thus achieving significant speedup.  \textsf{IND-generation} is also expensive for all methods, but it is a  standard step and the latency is comparable across all methods. In \textsf{Global-Predict}, we see that \abi (using the $k$-MCA algorithm) is highly efficient here. In comparison, baselines like MC-FK and HoPF are very expensive. 

\textbf{Efficiency of k-MCA-CC.} Our key innovation in \abi is the $k$-MCA-CC formulation and its efficient solutions. Despite its hardness,  we solve it optimally  (Algorithm~\ref{alg:k-mca-cc}), which is efficient as shown by  the purple \textsf{Global-Predict} component in Figure~\ref{fig:breakdown}. 


 
To drill down and study the latency of this key step, Figure~\ref{fig:k-mca-time-1000} shows  the  distribution of latency when we solve k-MCA-CC on the 1000-case  \textsc{Real}. We can see that the latency for the vast majority of cases is sub-second. In fact, the 50-th, 90-th and 95-th percentile latency  are 0.02, 0.06, and 0.17 seconds, respectively. This confirms that our Algorithm~\ref{alg:k-mca-cc} is not only optimal but also real-time and practical. 
There are a few cases where the latency is over 1 second, which are mostly cases with more than 40 input tables, which we argue is acceptable considering the size of the data. We also report that the largest case we encounter among all 10K+ BI models has 88 input tables, which  k-MCA-CC still solves optimally in about 11 seconds (not shown in the figure as it is not sampled in the 1000-case \textsc{Real}). This confirms that our algorithm is practical in solving large real-world BI cases both optimally and efficiently.



Figure~\ref{fig:efficiency_ablation} shows the benefit of the two key optimization techniques we used in solving k-MCA-CC: (1) artificial root in $k$-MCA (Algorithm~\ref{alg:k-mca}) and (2) branch-and-bound in $k$-MCA-CC (Algorithm~\ref{alg:k-mca-cc}). We  compare the  number of 1-MCA calls with and without the optimizations. (Note that we use the number of 1-MCA calls instead of wall-clock time, because the algorithm can timeout without our optimizations). 
We can see that the two optimization techniques give 5 and 4 orders of magnitude improvement in efficiency, respectively (the y-axis is in log-scale). Compared to a brute-force approach that solves k-MCA without optimization, the combined benefit is a staggering 10 orders of magnitude,  again showing the importance of our algorithms.

\vspace{-3mm}
\subsection{Ablation Studies}
We perform an ablation study to analyze the benefits of various \abi components. Our results are summarized in Figure~\ref{fig:ablation}.

\noindent\textbf{The effect of FK-once constraint.} The bar marked as  ``\textsf{no-FK-once-constraint}'' in Figure~\ref{fig:ablation} shows the result quality that removes the FK-once constraint (or using k-MCA instead of k-MCA-CC). 
There is a noticeable drop in precision/recall, showing the importance of this constraint in k-MCA-CC.

\noindent\textbf{The effect of precision-mode.} The  ``\textsf{no-precision-mode}'' bar shows the result quality if we omit the precision-mode step (Section~\ref{subsec:abi-p}) and use the recall-mode directly (Section~\ref{subsec:abi-r}). We see a substantial drop in precision (6 and 13 percentage points at edge-level and case-level, respectively), again showing the importance of k-MCA-CC.


\noindent\textbf{The effect of N-1/1-1 separation in local classifier.} The ``\textsf{no-N-1/1-1-seperation}'' bar shows the results when we use one local classifier, instead of splitting N-1 and 1-1 joins into two prediction tasks (Section~\ref{subsec:local-join}). The drop in quality is again evident.

\noindent\textbf{The effect of label transitivity in local classifier.} The bar for ``\textsf{no-label-transitivity}'' shows the drop in quality when we do not apply label transitivity in  local classification  (Section~\ref{subsec:local-join}). The effect is noticeable but diluted across 1000 cases, as the benefit  is more pronounced in large cases and less obvious in smaller ones. 

\noindent\textbf{The effect of no data-value features in local classifier.} The ``\textsf{no-data-features}'' bar shows the effect of removing features relating to data-value in our local classifier (we use meta-data features only instead). This directly corresponds to our lightweight \abi-S (Section~\ref{subsec:exp_overall}). The drop in quality is apparent but not too significant, which gives us a useful trade-off between quality and latency. 


\noindent\textbf{The effect of using local classifier only.} 
Lastly, we perform an important ablation study, to understand the benefit of our graph-based k-MCA algorithm on top of local classifier results.
The ``\textsf{LC-only}'' bar shows the performance if we  employ the same local classifier at the edge-level (keeping edges that have calibrated probability over 0.5), without using a holistic k-MCA optimization. The difference in performance is significant (25 percentage points in case-precision), underscoring the importance of our graph-based k-MCA.

\iftoggle{fullversion}
{
\subsection{Sensitivity Analysis}
Figure~\ref{fig:sensitivity} shows a  sensitivity analysis of the two parameters in \abi, the penalty-term $p$ in $k$-MCA (Section~\ref{subsec:abi-p}), and the edge-weight threshold $\tau$ in EMS (Section~\ref{subsec:abi-r}). Recall that because we use calibrated true probabilities,  $0.5$ is the natural choice in both cases.  This is confirmed in our study. Figure~\ref{fig:sen1} shows that when varying $p$ from 0 to 1, the region around 0.5 indeed gives the best result.
Figure~\ref{fig:sen2} shows the sensitivity to $\tau$, which is used to prune unpromising edges, where  a lower $\tau$ naturally leads to better recall at the cost of precision. We see that $\tau=0.5$ strikes a reasonable balance between precision/recall and produces high F-1, which is also our default setting in \abi.

\begin{figure}
\centering
\subfigure[Sensitivity to $p$ in $k$-MCA-CC]{
\label{fig:sen1}
\includegraphics[width=0.23\textwidth]{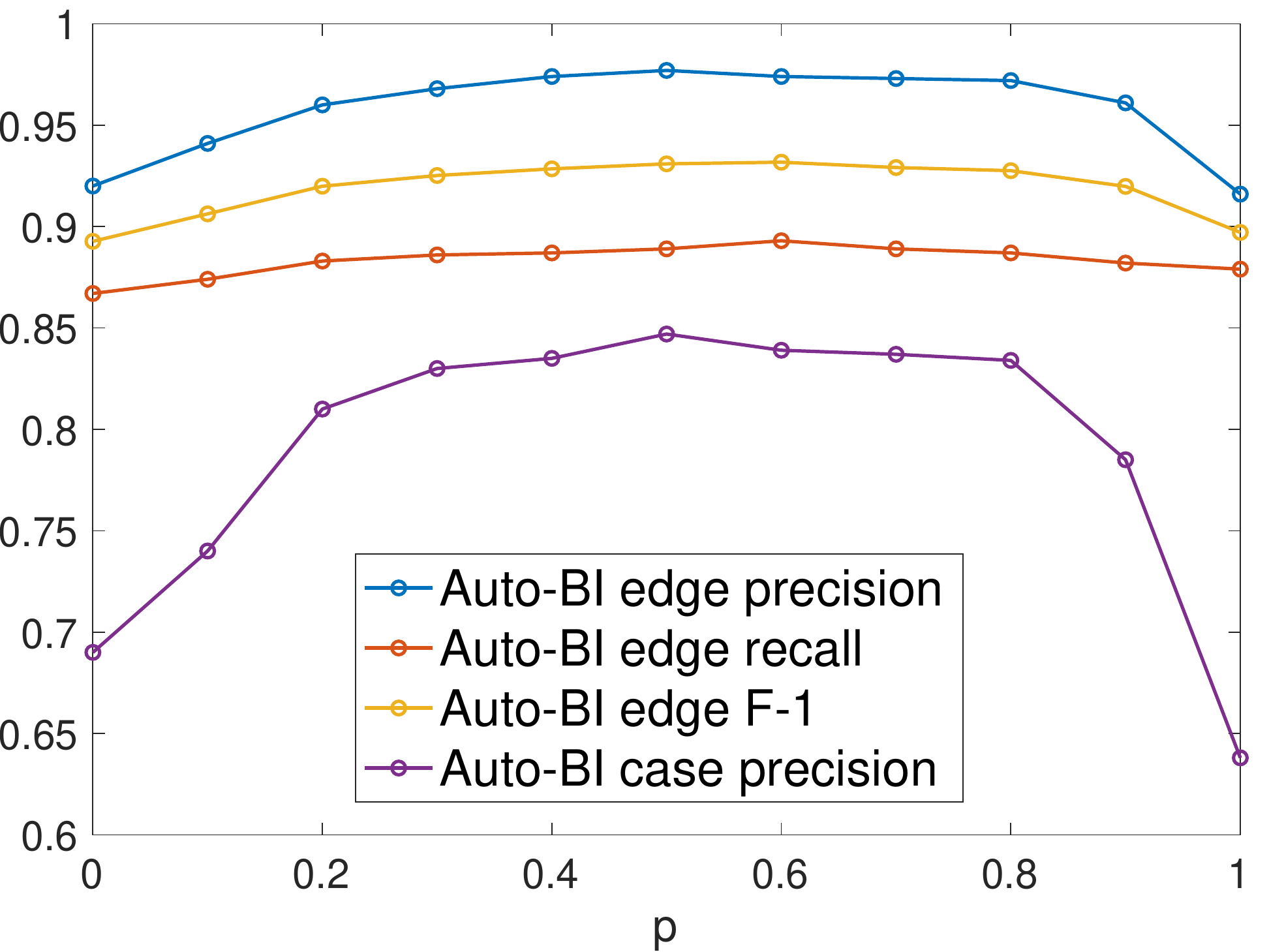}}
\subfigure[Sensitivity to $\tau$ in EMS]{
\label{fig:sen2}
\includegraphics[width=0.23\textwidth]{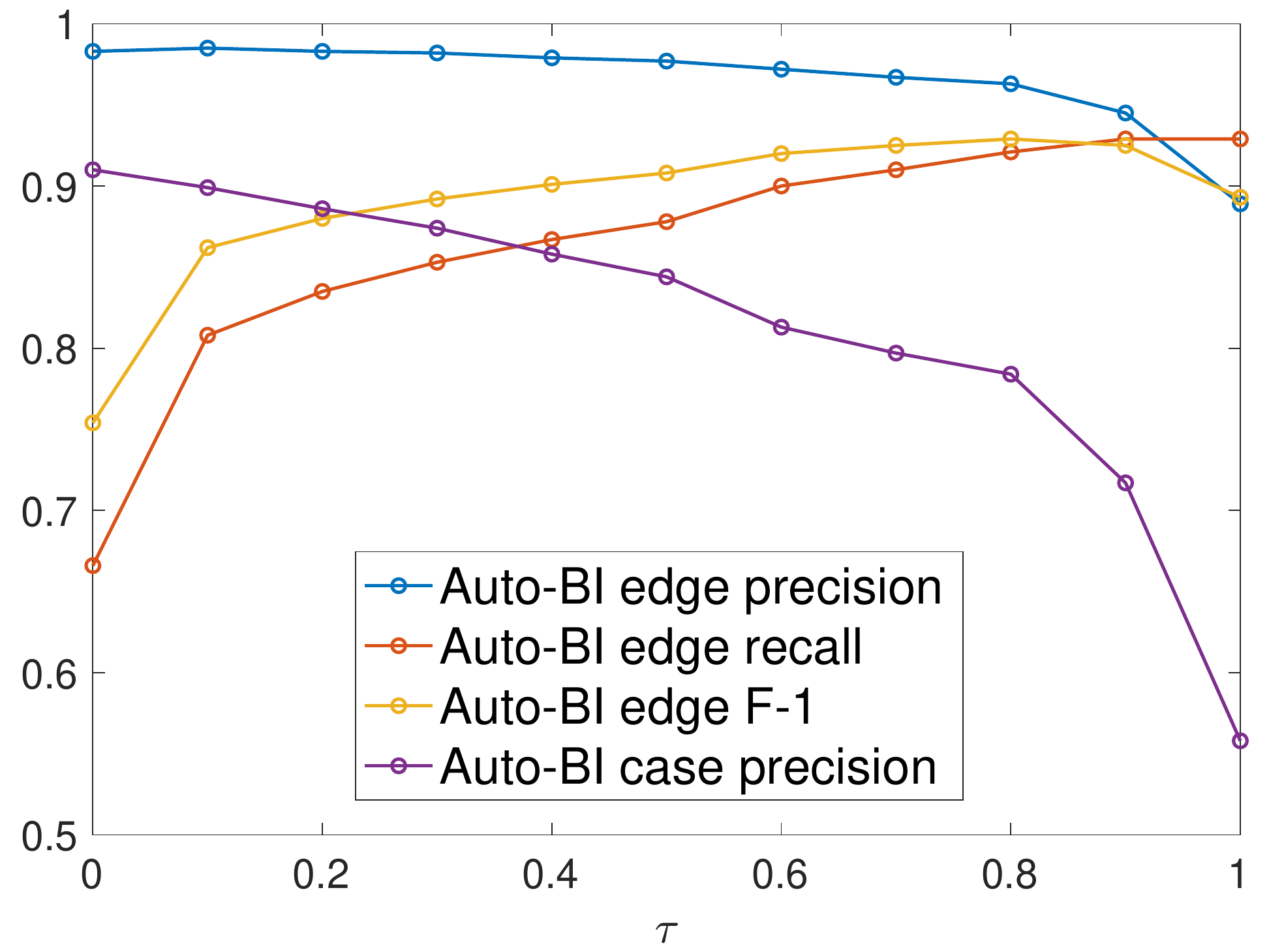}}
\caption{Sensitivity study on \textsc{Real} benchmark. }
\label{fig:sensitivity}
\end{figure}
}
{
\underline{Additional experiment results.} We present additional results, such as sensitivity analysis, as well as results on additional synthetic benchmarks (FoodMart, Northwind, AdventureWorks and WorldWide-Importers), in a full version of our paper~\cite{full} in the interest of space.
}
\section{Conclusions and Future Work}
In this work we develop an \abi system that can accurately predict join relationships in BI models, leveraging a novel graph-theoretical formulation called k-MCA that exploits snowflake-like structures common in BI models. 

Future directions include  improving the end-to-end efficiency of \abi, 
and extending the system to synthesize transformation steps, in order to automate BI model building end-to-end.

\clearpage

\bibliographystyle{ACM-Reference-Format}
\bibliography{auto-bi}

\clearpage

\iftoggle{fullversion}
{
    \appendix
    \newpage

\section{Local join: Two Optimizations}
\label{apx:local-join-two-opt}

\textbf{Separate N-1 and 1-1 classifiers.}
Unlike key/foreign-key that are typically N-1 joins, 
joins in real BI models can also be 1-1 joins, especially between related dimension tables describing the same logical entity (e.g.,  ``\code{employee-salaries}'' and ``\code{employee-details}''). 

Initially, we use one generic classifier to predict joinability between $(C_i, C_j)$ regardless of 1:1 or N:1. While performing an error analysis on incorrect predictions, we realized that while both 1:1 or N:1 are joins, they represent  conceptually different relationships with disparate characteristics. For example, 1:1 joins tend to be between the primary-keys of two dimension tables on the same logical entity (e.g. ``\code{employees}''), often with the same number of rows from both tables, producing perfect 1-to-1 matches (N:1 joins are, on the other hand, very different). Furthermore, because  1:1 joins tend to be between tables about the same logical entity with complementary information,  the column-headers for both tables that do not join will also have high overlap (e.g., with the prefix ''\code{employees\_}'' in the example of  ``\code{employee-salaries}'' and ``\code{employee-details}''). 

We thus treat N:1 and 1:1 as two separate classification tasks. This separation of is  unique in \abi and we will show the effect of this in our ablation studies.

\textbf{Enhance labels with transitivity.}
Given that we deal with entire BI models that have multiple tables, using the pair-wise join/not-join labels is no longer sufficient, because joinability can be ``transitive''. 

Consider, for instance, a table ``\code{Fact\_Sales}''  has a FK column called ``\code{sales\_emp\_id}'', that N:1 joins with  table ``\code{Dim\_Employee}'' on its PK column ``\code{emp\_id}'', which in turn 1:1 joins with table ``\code{Dim\_Employee\_Salaries}''  also on its PK ``\code{emp\_id}''. In such cases, even though  ``\code{Fact\_Sales.sales\_emp\_id}'' does not directly join with ``\code{Dim\_Employee\_Salaries.emp\_id}'', and the pair would ordinarily be considered a negative example (should not join), the two actually refer to the same semantic concept (employee\_ids) and are logically joinable. 
Formally, given the joinability label $L_{ij}=1$ for $(C_i, C_j)$ and $L_{jk}=1$ for $(C_j, C_k)$, we mark $L_{ik}=1$ for $(C_i, C_k)$ in training, even if there is no explicit join between $(C_i, C_k)$. Applying label transitivity overcomes incomplete ground-truth and makes it easier for ML models to fit the ground-truth, which leads to better quality end-to-end.

We note that the transitivity-based optimization is unique when we deal with general types of joins on a large graph, which is not needed when only pair-wise joins are considered, or when only PK-FK (N:1) joins are used as they are typically not transitive. 
\section{Local join classifier: Details}
\label{apx:features}

We give details on the Local Classifier (LC) in Section~\ref{subsec:local-join}, which takes two columns $(C_i, C_j)$ and predicts the joinability label $L_{ij}$.

As discussed in Section~\ref{subsec:local-join}, we separate the prediction problem for N:1 and 1:1 joins, since the two types of joins are  different at a conceptual level. For each classifier, we featurize each column pair $(C_i, C_j)$, which can be categorized as metadata-features (column-names, table-names) as well as actual data-features (e.g., data value overlap). In the following, we will describe the features of both the N:1  and 1:1 classifiers.

\textbf{N:1  classifier.} We design the following features for the N:1 classifier. We list the name of each feature, as well as its description.

\underline{Metadata-features.} These are features relating to column-names and table-names. We pre-process all column/table names, by first standardizing all names into tokens, based on camel-casing and delimiters (e.g., dash or underscores). 

\begin{itemize}[noitemsep,topsep=0pt,parsep=0pt,partopsep=0pt,leftmargin=*]
\item \textit{Jaccard\_similarity, Jaccard\_containment, Edit\_distance, \\ Jaro\_winkler}: These are standard string similarity between names. We compute these similarities between two column names $(C_i, C_j)$. In addition, assuming $C_i$ is the ``$1$'' side, we also compute the similarities between $(T_i + C_i, C_j)$, where $T_i$ is the name of the table from the ``1'' side, where the observation is that while column-names from fact-tables are often fully descriptive (e.g., ``\code{Employee-ID}''), sometimes column-names from dimension tables are simplified (e.g., ``\code{ID}'' and `\code{Name}''),  with the central entity (e.g., ``\code{Employees}'') only mentioned in table-names, thus making it necessary to piece together table-names and column-names to recover complete metadata. We use $max(sim(C_i, C_j), sim(T_i + C_i, C_j))$ as features for metadata similarity (where $sim$ can be different forms of similarity functions mentioned above, for a total of 4 features).

\item \textit{Embedding\_similarity}: In addition to string similarity, we also use embedding similarity, specifically SentenceBERT~\cite{SentenceTransformers,reimers2019sentence} trained on top of \textsf{all-mpnet-base-v2}), to compute the column name similarity, using a setup similar to above. 

\item \textit{Token\_count, Char\_count}: the number of tokens and characters in column names, for both $C_i, C_j$. These serve as auxiliary features on top of similarity scores above (e.g., longer column-names with high similarity may be a more reliable indicator of match, compared to shorter ones with the same similarity scores).
\item \textit{Col\_frequency}: the frequency of the column name $C_i$ (and $C_j$) in all BI-models we collect in the training set. Intuitively, matches of common names (e.g., \code{Code} and \code{Index}) are less reliable, so this feature works similarly to Inverse-Document-Frequency (IDF) in TF-IDF. 
\item \textit{Col\_position}: This feature keeps the positional index  of $C_i$ and $C_j$, counting from left (e.g., the 2nd column from left). This is based on the observation that columns on the left are more likely to join. 
\item \textit{Col\_relative\_position}: This is the same feature as \textit{Col\_position} above, except that it is measured in  relative terms, defined as: $\frac{\text{Col\_position}}{\text{Num\_of\_total\_columns}}$. 
\item \textit{Unique\_col\_position}: Similar to \textit{Col\_position}, except here we count unique columns, where the observation is that the first few \textit{unique} column from the left of a table is more likely to be PK for join. 
\end{itemize}

\underline{Data-features}. These features relate to the actual data content in columns, such as value overlap. Data-features provide complementary signals to  metadata-features above,  but are generally more expensive to process.

\begin{itemize}[noitemsep,topsep=0pt,parsep=0pt,partopsep=0pt,leftmargin=*]
\item \textit{Left\_containment, Right\_containment, Max\_containment}: value containment is an important signal for PK/FK joins~\cite{zhang2010multi, chen2014fast, hpi}. we compute  containment in both directions (left and right), and also the max of the two, for a total of three features.

\item \textit{Value\_distinct\_ratio}: This feature calculates the distinctness of columns, or the fraction of values in $C_i$ and $C_j$ that are distinct. 

\item \textit{Range\_overlap}: computes the overlap of the min/max value ranges, if both $C_i$ and $C_j$ are of numeric types.

\item \textit{EMD\_score}: This is a feature based on distributional-similarity and proposed in \cite{zhang2010multi}, we use this to complement the overlap-based features above.

\item \textit{Value\_length}: we cast all values in $C_i$ and $C_j$ to string type, and compute the average value length (longer values tend to produce more reliable matches).

\item \textit{Value\_type}: We one-hot encode column value types, such as integer, float, string.

\item \textit{Row\_cnt}: We use the number of rows in both tables as an auxiliary feature, where the observation is that fact table tends to have more rows compared to \textsf{dimension} tables.
  
\item \textit{Row\_ratio, Col\_ratio, Cell\_ratio}: These use similar intuition as \textit{Row\_cnt} above, except that we featurize the ratio of the number of rows/cols/cells explicitly between two tables. 
\end{itemize}

\underline{Feature importance.} Our results suggest that, for the N:1 classifier, the following features are the most important (in that order, based on sklearn output): \textit{Max$\_$containment}, \textit{Jaccard\_similarity}, \textit{Col\_relative\_position}, \textit{Edit\_distance,  Jaro\_winkler}, \textit{Range$\_$overlap}, \textit{EMD$\_$score}, \textit{Embedding$\_$similarity}, 
\textit{Col$\_$frequency}. 


\textbf{1:1  classifier.} Our 1:1 classifier share many of the same features as the N:1 classifier (except the \textit{Row\_ratio, Col\_ratio, Cell\_ratio} features, which are applicable to N:1 fact-dimension joins, but would be as useful for 1:1 joins). 
We omit these identical features, and 
 only describe features that are unique to the 1:1 classifier below.

\underline{Metadata-features}. These are features based only on column-names and table-names, like in the N:1 classifier.

\begin{itemize}[noitemsep,topsep=0pt,parsep=0pt,partopsep=0pt,leftmargin=*]
\item \textit{Table\_embedding}: We measure the SentenceBERT embedding similarity of table names where $T_i$ and $T_j$, based on the intuition that two tables with 1-1 join should likely refer to the same entity (e.g., \textsf{Employees} and \textsf{Employee-Details}, or \textsf{Country} and \textsf{Country-Code}, etc.), making high table-name similarity a useful signal.

\item \textit{Header\_jaccard}: measures the jaccard similarity between  all column-names of $T_i$ and $T_j$. This is based on our observation that two overlapping fact tables with highly similar column names do not 1:1 join in BI models, since such joins produce mostly redundant information.  Higher \textit{Header\_jaccard} is thus inversely correlated to joinability.
\end{itemize}

\underline{Data-features}. These are features relating to column-values.

\begin{itemize}[noitemsep,topsep=0pt,parsep=0pt,partopsep=0pt,leftmargin=*]

\item \textit{Min\_containment}: Instead of using \textit{Max\_containment}  between the \textit{Left\_containment} and \textit{Right\_containment} as in the N:1 classifer above, we use \textit{Min\_containment} for as the feature 1:1, based on the observation that 1:1 joins tend to join tuple-for-tuple between two tables, unlikely PK/FK joins.

\end{itemize}

\underline{Feature importance.} Our results suggest that for the 1:1 classifier, the following feature are the most important (in that order, based on sklearn output):
\textit{Min$\_$containment}, 
\textit{Col\_position}, 
\textit{Jaccard\_similarity}, 
\textit{Col\_relative\_position}, 
\textit{EMD$\_$score}, 
\textit{Header$\_$jaccard}, 
\textit{Col$\_$frequency} and \textit{Embedding\_similarity}.  


\begin{table}[]
\scriptsize
    \centering
    \begin{tabular}{|c|l|c|c|c|c|}
    \hline
    & Methods & Average & 50$\%$tile & 90$\%$tile & 95$\%$tile \\ \hline
    \multirow{3}{4em}{\abi}&\abi-P & 2.21 & 0.61 & 4.91 & 10.36 \\ 
    &\abi & 2.23 & 0.61 & 4.92 & 10.38   \\ 
    &\abi-S & 0.45 & 0.09 & 2.02 & 2.65 \\ \hline
    \multirow{3}{4em}{Original Baselines}& MC-FK & 1.21 & 0.33 & 2.79 & 3.59 \\ 
    &Fast-FK & 0.56 & 0.08 & 1.84 & 2.93 \\ 
    &HoPF & 4.31 & 0.25 & 6.74 & 18.72 \\ \hline
    \multirow{3}{4em}{Enhanced Baselines}&MC-FK+LC & 2.13 & 0.60 & 4.91 & 10.34 \\ 
    &Fast-FK+LC & 2.15 & 0.61 & 4.92 & 10.36\\ 
    &HoPF+LC & 6.41 & 1.13 & 11.88 & 27.23 \\ \hline
    \end{tabular}
    \caption{Comparison of end-to-end latency. Enhanced baselines (+LC) pay the cost of classifiers, which have latency comparable to \abi methods.}
    \label{tab:appendix_end_to_end_time}
    
\end{table}

\section{Additional Experiments}
\label{apx:add-exp}

\begin{table*}[]
\vspace{-5mm}
\scriptsize
    \centering
    \scalebox{0.92}{
    \begin{tabular}{|c|l||c|c|c|c||c|c|c||c|c|c||c|c|c||c|c|c||}
    \hline
    & & \multicolumn{4}{c||}{Real (OLAP)} & \multicolumn{6}{c||}{Synthetic (OLAP)} & \multicolumn{6}{c||}{Synthetic (OLTP)}  \\ \hline
     &Benchmark &\multicolumn{4}{c||}{1000-case \textsc{Real} benchmark} & \multicolumn{3}{c||}{TPC-H} & \multicolumn{3}{c||}{TPC-DS} & \multicolumn{3}{c||}{TPC-C} & \multicolumn{3}{c||}{TPC-E}\\ \hline
    & Metric & $P_{edge}$ & $R_{edge}$ &  $F_{edge}$ & $P_{case}$ & $P_{edge}$ & $R_{edge}$ &  $F_{edge}$  & $P_{edge}$ & $R_{edge}$ &  $F_{edge}$ & $P_{edge}$ & $R_{edge}$ &  $F_{edge}$ & $P_{edge}$ & $R_{edge}$ &  $F_{edge}$\\ \hline
    \multirow{3}{4em}{\abi}&\abi-P & \textbf{0.98} & 0.664 &  0.752 & \textbf{0.92} & \textbf{1} & 0.88& 0.93  & \textbf{0.99} & 0.28& 0.43 & \textbf{1} & 0.6& 0.75 & \textbf{1} & 0.4& 0.58\\ 
    &\abi & \textbf{0.973} &  \textbf{0.879} &  \textbf{0.907} & 0.853 & \textbf{1} & \textbf{1}& \textbf{1} & 0.96 & \textbf{0.91}& \textbf{0.93} & \textbf{1} & \textbf{0.8}& \textbf{0.89}& 0.96 & \textbf{0.93}& \textbf{0.95}\\
    &\abi-S & 0.951 &  0.848 &  0.861 & 0.779 & \textbf{1} & \textbf{1} & \textbf{1} & 0.92 & 0.89 & 0.91 & \textbf{1} & 0.7 & 0.82& 0.93 & \textbf{0.94} & \textbf{0.94}\\ \hline
    Commercial & \pbi & 0.916 & 0.584 & 0.66 & 0.754 &0 & 0 & 0 &0 & 0 & 0& 0 & 0 & 0 & 0 & 0 & 0 \\ 
    \hline
    \multirow{3}{4em}{Original Baselines}& MC-FK & 0.604 & 0.616 & 0.503 & 0.289 & \textbf{1} & \textbf{1}&\textbf{1}  & 0.73 & 0.65 & 0.68 & 0.46 & \textbf{0.8} & 0.63 & 0.57 & 0.79 & 0.48 \\ 
    &Fast-FK & 0.647 & 0.585 & 0.594 & 0.259 & 0.71 & 0.88 & 0.79& 0.62 & 0.35 & 0.44 & 0.62 & 0.57 & 0.6& 0.73 & 0.84 & 0.78\\ 
    &HoPF & 0.684 & 0.714 & 0.67 & 0.301 &0.86 &0.75 &0.8&0.87 &0.51 &0.65 &0.75 &0.7 &0.72 &0.71 &0.91 &0.81\\ 
    \hline
    \multirow{4}{4em}{Enhanced Baselines}&MC-FK+LC & 0.903 & \textbf{0.872} & 0.887 & 0.636 & \textbf{1}  & \textbf{1} & \textbf{1}  & 0.89 & 0.87 & 0.88 & 0.56  & \textbf{1} & 0.57& 0.92  & 0.83 & 0.88\\ 
    &Fast-FK+LC & 0.898 &  \textbf{0.879} &  0.883 & 0.631 & \textbf{1}&0.88 & 0.93 & 0.94&0.5 & 0.6 &\textbf{1}&0.7 & 0.82& 0.94 &0.87 & 0.91\\ 
    &HoPF+LC & 0.738 & 0.765 & 0.726 & 0.524 & \textbf{1} & 0.88 & 0.93 & 0.93 & 0.53 & 0.68 & \textbf{1} & 0.7 & 0.82& 0.91 & 0.88 & 0.9\\ 
    &LC & 0.885 & 0.864 & 0.87 & 0.631 & \textbf{1} & 0.88 &  0.93 & 0.85 & \textbf{0.91} & 0.88 & \textbf{1} & 0.6 &  0.75& \textbf{1} & 0.6 &  0.75\\ 
    \hline
    \end{tabular}}
    \caption{Quality comparison on the 1000-case \textsc{Real} benchmark and 4 TPC benchmarks. }
    \label{tab:apx_quality}
\end{table*}

\begin{table*}[]
\scriptsize
    \centering
    \scalebox{0.89}{
    \begin{tabular}{|c|c|c|c|c|c|c|c|c|c|c|c|}
    \hline
    &\# of tables  & 4 & 5 & 6 & 7 & 8 & 9 & 10 & [11,15] & [16,20] & 21+ \\ \hline
\multirow{3}{4em}{\abi}&\abi &\textbf{0.97} (0.99,0.95)&\textbf{0.97} (0.98,0.96)&\textbf{0.96} (0.99,0.94)&\textbf{0.95} (0.98,0.91)&\textbf{0.95} (0.99,0.92)&\textbf{0.96} (1.00,0.93)&\textbf{0.94} (0.98,0.90)&\textbf{0.90} (0.97,0.85)&\textbf{0.84} (0.95,0.75)&\textbf{0.79} (0.94,0.69)\\ 
&\abi-P &0.91 (1.00,0.84)&0.91 (0.99,0.85)&0.88 (1.00,0.79)&0.81 (0.98,0.69)&0.83 (0.98,0.71)&0.83 (1.00,0.70)&0.81 (0.99,0.69)&0.71 (0.98,0.56)&0.60 (0.96,0.44)&0.55 (0.95,0.39)\\ 
&\abi-S & 0.95 (0.98,0.93)&0.95 (0.96,0.94)&0.94 (0.97,0.92)&0.94 (0.97,0.90)&0.95 (0.98,0.91)&0.93 (0.97,0.89)&0.92 (0.96,0.88)&0.85 (0.92,0.79)&0.78 (0.90,0.70)&0.74 (0.89,0.63)\\ \hline
Commercial & \pbi &0.76 (0.94,0.66)&0.67 (0.91,0.55)&0.76 (0.94,0.66)&0.76 (0.93,0.66)&0.75 (0.91,0.65)&0.78 (0.91,0.70)&0.77 (0.92,0.67)&0.74 (0.90,0.65)&0.65 (0.80,0.56)&0.66 (0.88,0.54)\\ \hline
\multirow{3}{4em}{Baselines}& MC-FK &0.69 (0.88,0.57)&0.65 (0.93,0.49)&0.63 (0.70,0.58)&0.65 (0.67,0.63)&0.62 (0.70,0.56)&0.56 (0.46,0.72)&0.54 (0.48,0.63)&0.54 (0.49,0.61)&0.52 (0.42,0.69)&0.42 (0.30,0.68)\\ 
&Fast-FK &0.76 (0.79,0.72)&0.76 (0.79,0.74)&0.68 (0.69,0.66)&0.53 (0.53,0.52)&0.65 (0.67,0.63)&0.47 (0.46,0.47)&0.45 (0.48,0.42)&0.49 (0.53,0.46)&0.47 (0.49,0.46)&0.42 (0.40,0.44)\\ 
&HoPF &0.83 (0.86,0.81)&0.77 (0.77,0.76)&0.72 (0.73,0.71)&0.63 (0.58,0.70)&0.68 (0.67,0.70)&0.55 (0.49,0.64)&0.62 (0.55,0.70)&0.61 (0.58,0.64)&0.57 (0.55,0.60)&0.49 (0.44,0.56)\\ \hline
\multirow{4}{4em}{Enhanced Baselines} & MC-FK+LC &0.86 (0.90,0.83)&0.85 (0.91,0.80)&0.87 (0.87,0.86)&0.89 (0.87,0.90)&0.88 (0.85,0.91)&0.90 (0.87,0.92)&0.87 (0.83,0.91)&0.84 (0.81,0.87)&0.78 (0.79,0.77)&0.74 (0.73,0.74)\\ 
&Fast-FK+LC &0.90 (0.89,0.90)&0.91 (0.91,0.92)&0.87 (0.87,0.88)&0.85 (0.85,0.85)&0.87 (0.88,0.86)&0.86 (0.87,0.84)&0.85 (0.85,0.85)&0.83 (0.82,0.84)&0.76 (0.73,0.78)&0.71 (0.68,0.75)\\ 
&HoPF+LC &0.85 (0.87,0.83)&0.81 (0.82,0.81)&0.77 (0.78,0.75)&0.70 (0.65,0.75)&0.75 (0.74,0.76)&0.62 (0.57,0.67)&0.64 (0.58,0.72)&0.70 (0.67,0.72)&0.64 (0.60,0.69)&0.53 (0.46,0.62)\\ 
&LC &0.89 (0.90,0.88)&0.89 (0.90,0.88)&0.87 (0.87,0.87)&0.88 (0.89,0.87)&0.86 (0.87,0.86)&0.89 (0.91,0.87)&0.86 (0.87,0.84)&0.83 (0.84,0.82)&0.77 (0.79,0.75)&0.72 (0.72,0.71)\\ \hline
    \end{tabular}
    }
    \caption{Edge-level quality reported as ``F-1 (precision, recall)'',  by number of input tables in \textsc{Real} benchmark.}
    \label{tab:appendix_pr_result}
\end{table*}

\begin{table}[]
\small
    \centering
    \scalebox{0.65}{
    \begin{tabular}{|c|c|c|c|c|c|c|c|c|c|c|c|}
    \hline
    &\# of tables  & 4 & 5 & 6 & 7 & 8 & 9 & 10 & [11,15] & [16,20] & 21+ \\ \hline
\multirow{3}{4em}{\abi} & \abi &\textbf{1.00}&0.96&0.95&0.89&0.95&0.97&0.85&0.78&0.64&0.55\\    
&\abi-P &\textbf{1.00}&\textbf{0.98}&\textbf{0.99}&\textbf{0.94}&\textbf{0.96}&\textbf{0.99}&\textbf{0.95}&\textbf{0.89}&\textbf{0.83}&\textbf{0.67}\\ 
&\abi-S & 0.99&0.95&0.93&0.93&0.95&0.80&0.76&0.69&0.49&0.31\\ \hline
Commercial & \pbi &0.91&0.87&0.81&0.85&0.77&0.78&0.74&0.75&0.52&0.54\\ \hline
\multirow{3}{4em}{Baselines}& MC-FK &0.76&0.68&0.41&0.34&0.19&0.14&0.1&0.13&0.09&0.05\\ 
&Fast-FK &0.68&0.65&0.39&0.14&0.32&0.08&0.09&0.12&0.09&0.03\\ 
&HoPF &0.78&0.57&0.42&0.21&0.32&0.06&0.18&0.19&0.18&0.11\\ \hline
\multirow{4}{4em}{Enhanced Baselines}&
MC-FK+LC &0.89&0.92&0.75&0.62&0.67&0.70&0.63&0.53&0.35&0.31\\ 
&Fast-FK+LC &0.88&0.92&0.77&0.64&0.78&0.73&0.67&0.49&0.22&0.21\\ 
&HoPF+LC &0.87&0.74&0.66&0.62&0.57&0.49&0.48&0.34&0.28&0.19\\ 
&LC &0.87&0.85&0.72&0.66&0.68&0.81&0.65&0.53&0.28&0.26\\ \hline
    \end{tabular}
    }
    \caption{Case-level precision, by the number of input tables in \textsc{Real} benchmark.}
    \label{tab:appendix_pc_result}
\end{table}

We present additional experimental results in this section, mainly focusing on understanding the reason behind the quality advantage of \abi over baselines. 

At a high level, the advantage of \abi main comes from two sources: (1) the local-classifier step (Section~\ref{subsec:local-join}), that uses  data-driven ML scoring together with principled probability calibration, and (2) the graph-based k-MCA step for global optimization (Section~\ref{sec:abi}). In order to understand the contributions of these two sources, we ``enhance'' existing baselines, by injecting our local-classifier scores from step (1) into their algorithm as follows.

\textbf{Enhanced baselines: MC-FK+LC, Fast-FK+LC, HoPF+LC}. Since MC-FK, Fast-FK and HoPF are  not competitive on the \textsc{Real} benchmark, we inject our Local-Classifier (LC) scores for join-likelihood into these baselines, replacing their heuristic scores with our calibrated classifier scores. This leads to stronger baselines, and allows us to see the benefit attributable to better local classifer scores, vs. the global k-MCA algorithm. To differentiate from the original baselines, we use a "+LC" suffix for these enhanced baselines, which become MC-FK+LC, Fast-FK+LC, and HoPF+LC, respectively.

\noindent\textbf{Quality Comparisons.} Table~\ref{tab:apx_quality} shows the additional comparison with MC-FK+LC, Fast-FK+LC, HoPF+LC, on top of the results reported in Table~\ref{tab:quality}. 
As we can see, on the \textsc{Real} benchmark, the enhanced baselines improve substantially over the original baselines. However, all baselines still lag behind \abi, especially in terms of precision: the edge-level error rate of \abi is 5x smaller than the best baseline (2\% vs. 10\%), while the case-level error rate of \abi is 4x smaller (8\% vs. 34\%). Because all the enhanced baselines use the same LC classifiers as \abi, the precision benefit can be attributable to the global optimization in the k-MCA step, underscoring the importance of our graph-based formulation. On the \textsc{Synthetic} TPC benchmarks, we have similar observations consistent with the \textsc{Real} benchmark. With the help of the better scores from the local classifier, the enhanced baselines improve over the original baselines on both OLAP benchmarks (TPC-H and TPC-DS) and OLTP benchmarks. (TPC-C and TPC-E) But overall, \abi still gains the best performance.

Table~\ref{tab:appendix_pr_result} and Table~\ref{tab:appendix_pc_result} report more detailed numbers, bucketized based on the number of input tables. Similar trends are observed here, as we see \abi is still the best method overall in terms of quality, across the spectrum of large and small test cases. The advantage is the most significant in precision, especially on larger test cases, which tend to be more difficult to predict correctly. 

\textbf{Latency comparisons}.  As we can see from Table~\ref{tab:appendix_end_to_end_time}, introducing the LC classifier step into baselines increases their latency, though not substantially. The enhanced baselines have latency comparable to \abi methods, with the exception of HoPF+LC, which is more expensive in terms of latency.

    \section{Proof of Theorem 3}
\label{apx:proof-kmcacc-inapprox}
\begin{proof}
We show the hardness and inapproximability of $k$-MCA-CC. We will first show the hardness of the problem using a reduction from Hamilton path~\cite{hartmanis1982computers}. We then show inapproximability using a reduction from non-metric min-TSP~\cite{escoffier2006completeness}.

Recall that given a graph $G=(V, E)$, the Hamilton path problem looks for a path that visits each vertex exactly once. We reduce Hamilton path to $k$-MCA-CC. Given an instance of Hamilton path on $G=(V, E)$, we construct an instance of $k$-MCA-CC  as follows. We construct a new graph $G'=(V, E')$ on the same set of vertices $V$, where for each $v_i \in V$ we construct a table $T_i$ with a single column $C_i$ in  $k$-MCA-CC.  For each directed edge $e(v_i, v_j) \in E$, we construct a possible N:1 FK/PK join between $C_i$ of $T_i$ and $C_j$ of $T_j$ in  $k$-MCA-CC, which would correspondingly create an edge $e(v_i, v_j) \in E'$ in the graph representation for $k$-MCA-CC. Note that because we construct each table $T_i$ to have a single column $C_i$, and given the FK-once constraint (Equation~\eqref{eqn:kmcacc_fk_once}), when there are multiple edges pointing away from a vertex $v$ in $G'$, a solution to $k$-MCA-CC is forced to pick only one such edge, ensuring that the $k$-MCA returned by the solution contains either single paths, or isolated vertices (which are trivial forms of single paths).
Furthermore, in our constructed $k$-MCA-CC, we set all edge-weight to be unit weight, and set the penalty weight $p$ to be a large constant $|V|+1$. This large penalty weight ensures that if a 1-MCA exists on $G'$, it is guaranteed to have a lower cost than any $k$-MCA with $k>1$, because the penalty term $|V|$ is already larger than the cost of 1-MCA (which is at most $|V|-1$). 

We now show that by this construction, a solution $P$ to Hamilton-path on $G$, will also be a solution to the $k$-MCA-CC we constructed. This is the case because if $P$ is a Hamilton-path of $G$,  it must be a feasible solution on $G'$ to $k$-MCA-CC, because first of all, $P$ is a single path, which is a trivial 1-MCA, satisfying Equation~\eqref{eqn:kmcacc_kmca}. Furthermore, since $P$ passes through each vertex exactly once, it satisfies our FK-once constraint in Equation~\eqref{eqn:kmcacc_fk_once}. Lastly, its cost in Equation~\eqref{eqn:kmcacc_obj} is exactly $|V|-1$ (with $|V|-1$ unit-weight edges in the path $P$), which is guaranteed to be lower than any $k$-MCA with $k>1$ (whose penalty cost alone is $|V|$). All of these above guarantee that $P$ is an optimal solution to the $k$-MCA-CC we constructed on $G'$. 

Assume for a moment that we can solve $k$-MCA-CC efficiently in polynomial time, then by this construction, we can solve any instance of Hamilton path using the reduction above, also in polynomial time, which contradicts with existing complexity result of Hamilton path~\cite{hartmanis1982computers}. We can thus conclude that  $k$-MCA-CC is also NP-hard. 

We now show the inapproximability of $k$-MCA-CC using a reduction from non-metric min-TSP~\cite{escoffier2006completeness}.
Recall that the non-metric min-TSP on a graph $G=(V, E)$ finds the min-cost cycle that visits each vertex exactly once, and returns to the origin, where cost is defined as the sum of edge-weights (which in the non-metric version of min-TSP, can be arbitrary numbers).

Recall that the Hamilton cycle problem (a special version of min-TSP) can be reduced to Hamilon path, by adding an artificial source and sink vertex $s$ and $t$ to the input graph, which is connected to a vertex $v$ and its cleaved copy $v'$ with the same neighborhood as $v$~\cite{hartmanis1982computers}. We perform a similar construction from non-metric TSP (cycle) to non-metric min-cost Hamilton path, using the same artificial source and sink vertex $s$ and $t$, together with a cleaved $v'$. For any instance of the non-metric min-TSP problem, we can thus construct a min-cost Hamilton path problem. Because the edges $(s, v), (t, v')$ are constructed to have 0 cost, the reduction is value preserving as an optimal solution to min-TSP is also the optimal min-cost Hamilton-path (modulo $s$ and $t$), and the two solutions have the same objective function values. Furthermore, using the same reduction from Hamilton path to $k$-MCA-CC above, we can then reduce an instance of non-metric min-TSP first to min-cost Hamilton path, and then to $k$-MCA-CC, all with the same objective values for optimal solutions. 

Given that the non-metric min-TSP is EXP-APX-complete~\cite{escoffier2006completeness}, we know $k$-MCA-CC is also EXP-APX-complete (for otherwise given any instance of non-metric min-TSP, we can use the reduction above to solve the corresponding $k$-MCA-CC efficiently, and thus non-metric min-TSP, contradicting with the complexity result of non-metric min-TSP).
\end{proof}
    \section{Proof of Theorem 4}
\label{apx:proof-optimality}

\begin{proof}
We show that Algorithm~\ref{alg:k-mca-cc} solves $k$-MCA-CC optimally. If the optimal solution to $k$-MCA, $J$, does not violate FK-once constraint (Equation~\eqref{eqn:kmcacc_fk_once}, then in Line~\ref{line:kmcacc-check} we know $J$ must be a feasible solution to $k$-MCA-CC on the same graph. Furthermore, because the feasible region of $k$-MCA is strictly no smaller than that of $k$-MCA-CC, we know if $J$ is an optimal solution to $k$-MCA and it is feasible  for $k$-MCA-CC, it must also be an optimal solution to $k$-MCA-CC. Thus we return $J$ in Line~\ref{line:kmcacc-return-early} and we are done.

If instead, the optimal solution to $k$-MCA, $J$, does violate FK-once constraint in Line~\ref{line:kmcacc-check}, because there are at least one edge set $C_s = \{ e_{sj}, e_{sk}, \ldots \} \subseteq J$ that causes violations, where all edges in $C_s$  point from the same vertex with the same column-index $s$. We partition $C_s$ into $|C_s|$ number of subsets $C_s^1$, $C_s^2$, $\ldots, C_s^{|C_s|}$, each with exactly one edge from $C_s$, and then construct $|C_s|$ number of $k$-MCA-CC problem instances, each with a new graph $G_i = (V, E_i)$, where $V_i = V$, $E_i = E \setminus C_s \cup C_s^i$, which we then recurse and solve the $k$-MCA-CC on each graph $G_i$ in Line~\ref{line:kmcacc-recurse}. To show that $J^* = \argmin_{J_i}{c(J_i)}$ is the optimal solution to the original $k$-MCA-cc problem on $G$, we need to show that the optimal solution to the original $k$-MCA-cc  problem on $G$, $J^+$, is still not pruned when we split edges in $C_s$ and create $|C_s|$ number of smaller problems with $G_i$. In order to see this, notice that the optimal solution to the original $k$-MCA-cc  problem on $G$, $J^+$, will have at most one edge in $C_s$ (otherwise it violates FK-once constraint and would not have been a feasible solution). If $J^+$ has no edge in $C_s$, then it has not been pruned away when we partition  $C_s$ and reduce the solution space, as it is still a feasible solution in each $G_i$. Alternatively, $J^+$ has one edge in $C_s$, and let it be the edge in $C_s^{l}$, then all the edges in $J^+$ is still in the graph $G_l$ (which only pruned edges in $C_s \setminus C_s^l$). As such, $J^+$ is still a feasible solution in  $G_l$, and will be returned in step Line~\ref{line:kmcacc-argmin}.
\end{proof}
}
{

}

\end{document}